\documentclass[twocolumn,a4paper]{aa}

\usepackage[suffix=]{epstopdf}
\begin{document}

\title{Diffusive Transport of Energetic Electrons in the Solar Corona: X-ray and Radio Diagnostics}
\author{S. Musset\inst{\ref{inst1},\ref{inst2}}
\and E. P. Kontar\inst{\ref{inst3}}
\and N. Vilmer\inst{\ref{inst1}}}
\institute{LESIA, Observatoire de Paris, PSL Research University, CNRS, Sorbonne Universit\'{e}s, UPMC Univ. Paris 06, Univ. Paris Diderot, Sorbonne Paris Cit\'{e}\label{inst1}
\and School of Physics and Astronomy, University of Minnesota\label{inst2}
\and School of Physics and Astronomy, University of Glasgow, Glasgow G12 8QQ, UK\label{inst3}}
\date{Received ... / Accepted ...}

\abstract {Imaging spectroscopy in X-rays with RHESSI provide the possibility to investigate the spatial evolution of the X-ray emitting electron distribution and therefore to study the transport effects on energetic electrons during solar flares.}
{We study the energy dependence of the energetic electron scattering mean free path in the solar corona.}
{We use the imaging spectroscopy technique with RHESSI to study the evolution of energetic electrons distribution in different part of the magnetic loop during the 2004 May 21 flare. These observations are compared with the radio observations of the gyrosynchrotron radiation of the same flare by \cite{kuznetsov_and_kontar_2015}, and with the predictions of the diffusive transport model described by \cite{kontar_et_al_2014}.}
{The X-ray analysis shows a trapping of energetic electrons in the corona and a spectral hardening of the energetic electron distribution between the top of the loop and the footpoints. Coronal trapping of electrons is stronger for the radio-emitting electrons than for the X-ray-emitting electrons. These observations can be explained by the diffusive transport model derived by \cite{kontar_et_al_2014}.}
{We show that the combination of X-ray and radio diagnostics is a powerful tool to study electron transport in the solar corona in different energy domains. We show that the diffusive transport model can explain our observations; and in the range 25-500 keV, the electron scattering mean free path decreases with electron energy. We can estimate for the first time the scattering mean free path dependence on energy in the corona. 
}


\titlerunning{Diffusive transport of energetic electrons in the solar corona}
\maketitle

%%%%%%%%%%%%%%%%%%%%%%%%
% *** INTRODUCTION *** %
%%%%%%%%%%%%%%%%%%%%%%%%

\section{Introduction}

Particle transport between the acceleration site and the X-ray and radio emission sites is a key process that must be studied and understood in order to use X-ray and radio diagnostics to study particle acceleration during solar flares. Indeed, transport mechanisms can modify the spatial and spectral distributions of energetic particles produced by the acceleration process.
The spatial and spectral distributions of X-ray emitting electrons can be studied during solar flares using imaging spectroscopy in X-rays. This technique is therefore a useful tool to study the transport of energetic electrons in magnetic loops.

In addition to imaging and spectroscopy of solar flares in X-ray and gamma-ray ranges \citep{rhessi}, the Reuven Ramaty High Energy Solar Spectroscopic Imager (RHESSI) provides the possibility to use imaging spectroscopy in hard X-rays (HXR). This technique has been used to study events which exhibit both footpoint and coronal HXR sources \citep[e.g.][]{krucker_and_lin_2002,emslie_et_al_2003,battaglia_and_benz_2006,piana_et_al_2007,simoes_and_kontar_2013}. 
These studies show in particular that in some events, X-ray emission in the coronal source is a combination of both thermal and non-thermal emissions.
\cite{battaglia_and_benz_2006} showed that the difference between the photon spectral indexes in the coronal source and the footpoint sources was between 1.2 and 0.6 (in three flares) and between 2.4 and 3.7 (in two flares), but not 2, the expected value in the standard model. This discrepancy between expected and observed differences implies that additionnal transport effects are needed to explain these observations. 
\cite{battaglia_and_benz_2006} interpreted the hardening of the spectrum as the result of a filter mechanism causing low-energetic electrons to preferentially loose energy before reaching the chromosphere; candidates for this mechanisms being collisions and the electric field of the return current.
More recently, \cite{simoes_and_kontar_2013} compared the electronic spectral indexes in coronal and footpoint sources and went to similar conclusions: on the four events studied, the difference of electronic spectral index between the footpoints and the coronal source lies between 0.2 and 1.0, while it is expected to be nul in the case of limited electron interaction with the ambiant medium during the transport in the loop. This study also shows that the rate of non-thermal electrons in the coronal source is larger than in the footpoints, by a factor ranging from $\approx$2 to $\approx$8. These observations suggest that a mechanism is responsible for energetic electron trapping in the coronal source. Such mechanism could be for instance magnetic mirroring or turbulent pitch-angle scattering. 
These observations carried with imaging spectroscopy in X-rays provide new constraints to electron propagation models and are not compatible with the predictions of the standard model, described in the following.

In the standard model of solar flares \citep[see e.g.][]{sturrock_1968,arnoldy_et_al_1968,sweet_1969,Syrovatskii_and_Shmeleva_1972}, energetic electrons are accelerated in the corona and then propagate along the magnetic field lines of coronal loops, losing a relatively small amount of their energy via collisions with the particles of the ambiant plasma, until they reach the chromosphere, a denser medium where they loose instantaneously the bulk of their energy and are thermalized. During their propagation, energetic electrons radiate a bremsstrahlung emission which is detected in the X-ray range, both in the coronal loop and in the footpoints \citep[see e.g.][for recent reviews]{holman_et_al_2011,kontar_et_al_2011_review}. 
In this standard model for the electron transport, we expect to see as many electrons leaving the looptop source than arriving in the footpoint, since the propagation time is much smaller than the collision time in the corona, and than the time cadence of X-ray observations. For that reason, it is also expected to see the same spectral distribution of energetic electrons in the looptop and in the footpoints. Therefore, in this standard model, we expect to find the same electron rate and the same electronic spectral index in the looptop and in the footpoints
However, as it has been described in this introduction, recent analysis of X-ray emission during solar flares \citep[e.g.][]{battaglia_and_benz_2006,simoes_and_kontar_2013} shown that this standard model for electron propagation could not explain their observations. 

Trapping of energetic electrons in the coronal part of the loop can be explained by the effect of a converging magnetic field. The simpliest way to modelize magnetic mirrors is to consider a magnetic loss cone for the pitch angle distribution. 
The value of the loss-cone angle depends on the magnetic ratio $\sigma = B_{FP}/B_{LT}$. 
\cite{aschwanden_et_al_1999,tomczak_and_ciborski_2007,simoes_and_kontar_2013} calculated the magnetic ratios needed to explain X-ray observations, assuming an isotropic pitch angle distribution, and found values lying between 1.1 and 5.0.  
However, magnetic loss cones are an approximation for magnetic mirroring only valid for rapid variations of density and magnetic field amplitude near the footpoints of the magnetic loop. More realistic models of magnetic convergence have been developed and the evolution of energetic electron populations in the case of a converging magnetic field have been studied analytically \citep[see e.g.][]{kennel_and_petschek_1966,kovalev_and_korolev_1981,leach_and_petrosian_1981,mackinnon_1991,melrose_and_brown_1976,vilmer_et_al_1986} or numerically \citep[see e.g.][]{bai_1982,mcclements_1992,siversky_and_zharvoka_2009,takakura_1986}.
These studies showed that the convergence of magnetic field causes energetic electron trapping in the corona, but the value of the ratio of electron rates in the corona and in the footpoints depends on numerous parameters such as the density, the form of the magnetic field convergence or the electronic pitch angle distribution. We note that \cite{takakura_1986} calculated in particular the difference of spectral index between the coronal source and the footpoints, lying between 0 and 0.8.

Energetic electron trapping can also be explained by an alternative scenario, the diffusive transport of electrons due to strong pitch angle scattering. Turbulent pitch angle scattering is the result of small scale magnetic fluctuations affecting the parallel transport of energetic electrons in flaring loops. The presence of such magnetic fluctuations is suggested by the increase of loop width which has been observed with RHESSI \citep{kontar_et_al_2011,bian_et_al_2011}. 
\cite{kontar_et_al_2014} studied the effect of strong turbulent pitch angle scattering, leading to a diffusive transport of energetic electrons in the loop, during solar flares. They compared the predictions of the model with observations of four flares and estimated for these events that the characteristic mean free path for this diffusive transport was of $10^{8}-10^{9}$ cm, which is smaller than the typical size of a loop ($\approx 2 \times 10^{9}$ cm) and comparable to the size of coronal sources ($\approx 5 \times 10^{8}$ cm). Therefore, the authors concluded that pitch-angle scattering du to magnetic fluctuations in a collisional plasma is likely to be present in flaring loops. 

The diffusive transport of electrons and ions is also studied since several decades in the interplanetary medium where in-situ measurements of particles are made. \cite{jokipii_1966} developed the first description of particle scattering in varying magnetic field. In this analysis, the magnetic field is considered as the superposition of a constant field and a smaller fluctuating component which is an homogeneous random function of position with zero mean. This work was improved in later approachs \cite[see e.g.][for a review]{droge_2000_rev}. Some studies focussed on the possible rigidity \footnote{The rigidity $R$ of a charged particle is defined by $R=pc/q$ where $p$ is the momentum of the particle, $q$ its charge and $c$ the speed of light. For relativistic particle, $R=\sqrt{E(E+2mc^2)}/q$ where $E$ is the kinetic energy and $m$ the mass of the particle. } dependence of the particle mean free path. \cite{palmer_1982} studied the values of the mean free path measured for solar particle events near the Earth and found that although the values could vary of two orders of magnitude, no dependence in rigidity was found: the values of the mean free path at different rigidities were found mostly between 0.08 and 0.3 AU, in the so-called 'consensus range'.
However, later studies revisited this consensus \citep[see e.g.][]{bieber_et_al_1994,droge_2000} and showed in particular that the electron scattering mean free path is rigidity dependant. In particular, \cite{droge_2000} showed that the electron mean free path varies as a power law with rigidity, in the range 0.1-1 MV, with a slope of -0.2. More recently, \cite{agueda_et_al_2014} found the same kind of rigidity dependence for six solar particle events (over seven studied), in the 0.3-0.5 MV range, with slopes varying between -0.3 and -1.2.
 
In this paper, we present X-ray observations of one flare which exhibit a non-thermal loop-top X-ray source.
The M2.6 flare on 2004 May 21 flare is located near the solar limb and was well observed by RHESSI, the Nobeyama Radio Heliograph (NoRH) and the Nobeyama Radio Polarimeters (NoRP). 
\cite{kuznetsov_and_kontar_2015} showed that the gyrosynchrotron emissions observed at 17 and 34 GHz with the NoRH were cospatial with the X-ray emission (even if the centroid of the X-ray emission is shifted of about 6 arcsec under the position of the centroid of the 34 GHz emission), where a looptop source and two footpoints are visible. They also deduced from the NoRP spectra of the microwave emission that the absolute value of the electronic spectral index was about 2.7.
The authors simulated the gyrosynchrotron emission with the recently developed IDL tool GX Simulator, using a linear force-free extrapolation of the magnetic field of the loop. The results of their simulation was compared with the microwave data to deduce the spatial and spectral properties of the radio-emitting energetic electrons. They found that the microwave emission is mostly produced by electrons of a few hundreds of keV having a hard spectrum (with an absolute value of the spectral index around 2). They also showed that the spatial distribution of energetic electrons with energy above 60 keV \footnote{the lower limit adopted in \cite{kuznetsov_and_kontar_2015} is 60 keV even if the radio emissivity is maximum for electrons of a few hundred of keV (Kuznetsov, private communication).} is strongly peaked near the top of the flaring loop, implying that there is a coronal trapping of energetic electrons during this event. The peak of the spatial distribution of energetic electrons is shifted of 3.2 Mm in regards to the top of the loop where the magnetic field is minimal.
According to the authors, this spatial distribution of energetic electrons is due to a combination of the processes of particle acceleration, trapping, and scattering. However, the authors did not calculate the scattering rate but focussed on the distribution of electrons in the loop.

The aim of this paper is the study of the trapping of energetic electrons in two distinct energy domains. For that purpose, we used the analysis of \cite{kuznetsov_and_kontar_2015} of the radio-emitting energetic electrons above a few hundred of keV, and we analysed the X-ray emission of energetic electrons with energies below 100 keV.
We therefore show in this paper that the electron scattering mean free path decreases with increasing electron energy.
In section \ref{observations} is presented the imaging spectroscopy of the 2007, May 21 flare in X-rays. The spatial and spectral distributions derived from the X-ray observations are presented in section \ref{section3}. The interpretation of the X-ray observations, the comparison between X-ray and radio observations, and the comparison with the predictions of the diffusive transport model of \cite{kontar_et_al_2014} are discussed in section \ref{interpretation}, along with the energy dependence of the energetic electron scattering mean free path in the frame of that model. Alternative mechanisms and improvement of the diffusive transport models are discussed in section \ref{discussion}. The main results are summarized in section \ref{conclusion}.

%%%%%%%%%%%%%%%%%%%%%%%%%%%%%%%%
% *** DATA AND METHODOLOGY *** %
%%%%%%%%%%%%%%%%%%%%%%%%%%%%%%%%

\section{X-ray imaging spectroscopy at the peak of the flare}
\label{observations}

\begin{figure}
\begin{center}
\includegraphics[width=\linewidth]{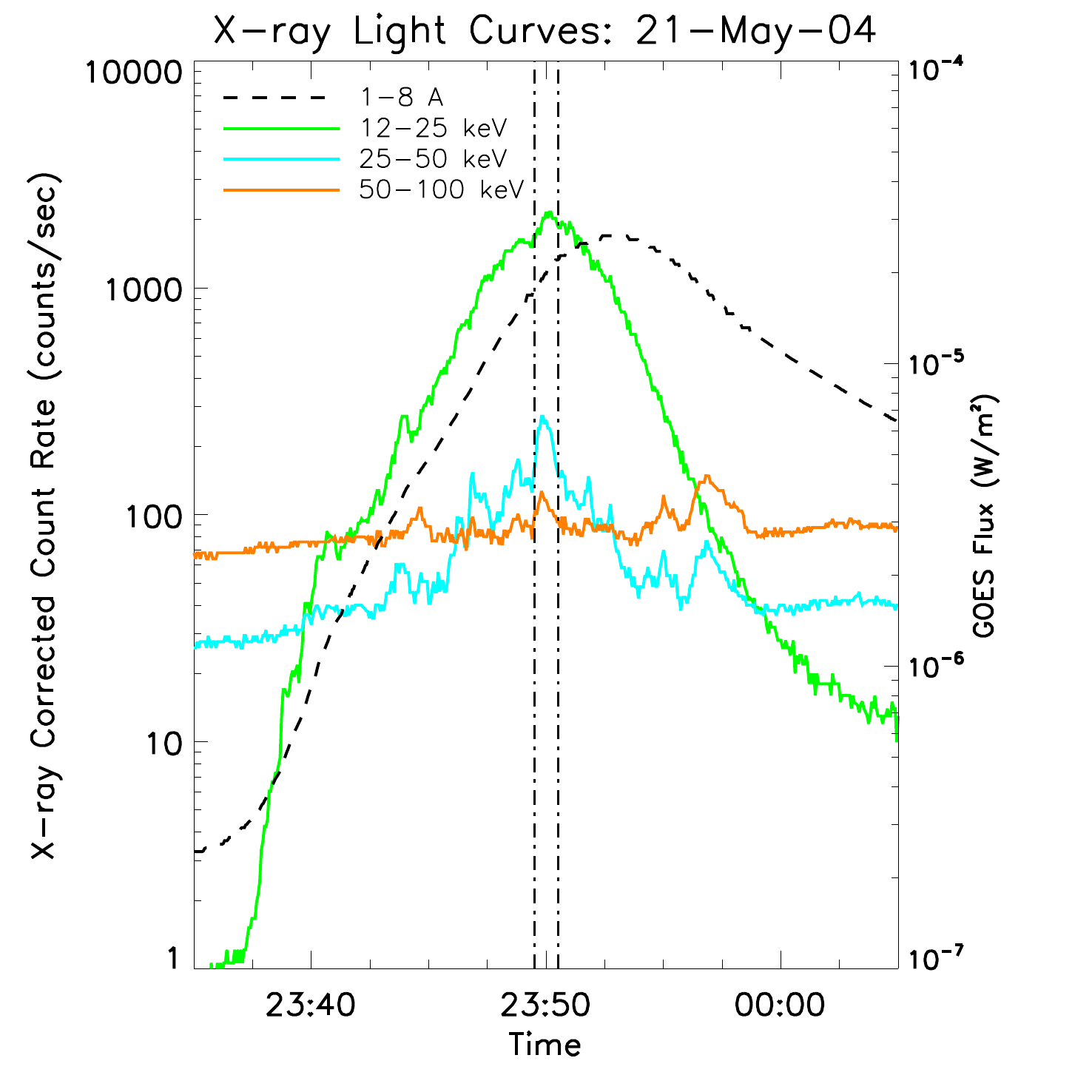}
\caption{RHESSI corrected count rates between 23:35 and 00:05 UT, in different energy ranges (green: 12-25 keV, cyan: 25-50 keV, orange: 50-100 keV) and GOES flux between 1.0 and 8.0 \AA \ (dashed line). The vertical dashed-dotted lines at 23:49:30 and 23:50:30 UT mark the time interval used for imaging spectroscopy.}
\label{lightcurves}
\end{center}
\end{figure}

The M2.6 flare on 2004 May 21 flare, in active region 10618, was detected by RHESSI in the 3-100 keV range. The RHESSI corrected count rates at relevant energies bins are presented on figure \ref{lightcurves}, together with the X-ray flux from GOES. In this figure the count rates are corrected from the changes of attenuator state and decimation state. 
The peak of the RHESSI count rates is around 23:50 UT, which is about 2 minutes before the GOES X-ray peak. On figure \ref{lightcurves}, the vertical dashed-dotted lines show the time interval (23:49:30 to 23:50:30 UT) chosen to image the X-ray emitting sources. We chose the time interval with the highest signal above 25 keV. Note that a consequent peak in the 25-100 keV lightcurve is visible between 23:56:00 and 23:58:30 UT; but it was not possible to reconstruct a reliable image above 25 keV during this time interval. 
The photon statistics is also too low to enable reliable imaging spectroscopy on other one-minute intervals after the X-ray peak of the flare, due to the high level of noise in the images, and therefore the time evolution of X-ray emission is not discussed in this paper.

\subsection{X-ray imaging of the source}

\begin{figure}
\begin{center}
\includegraphics[width=0.8\linewidth]{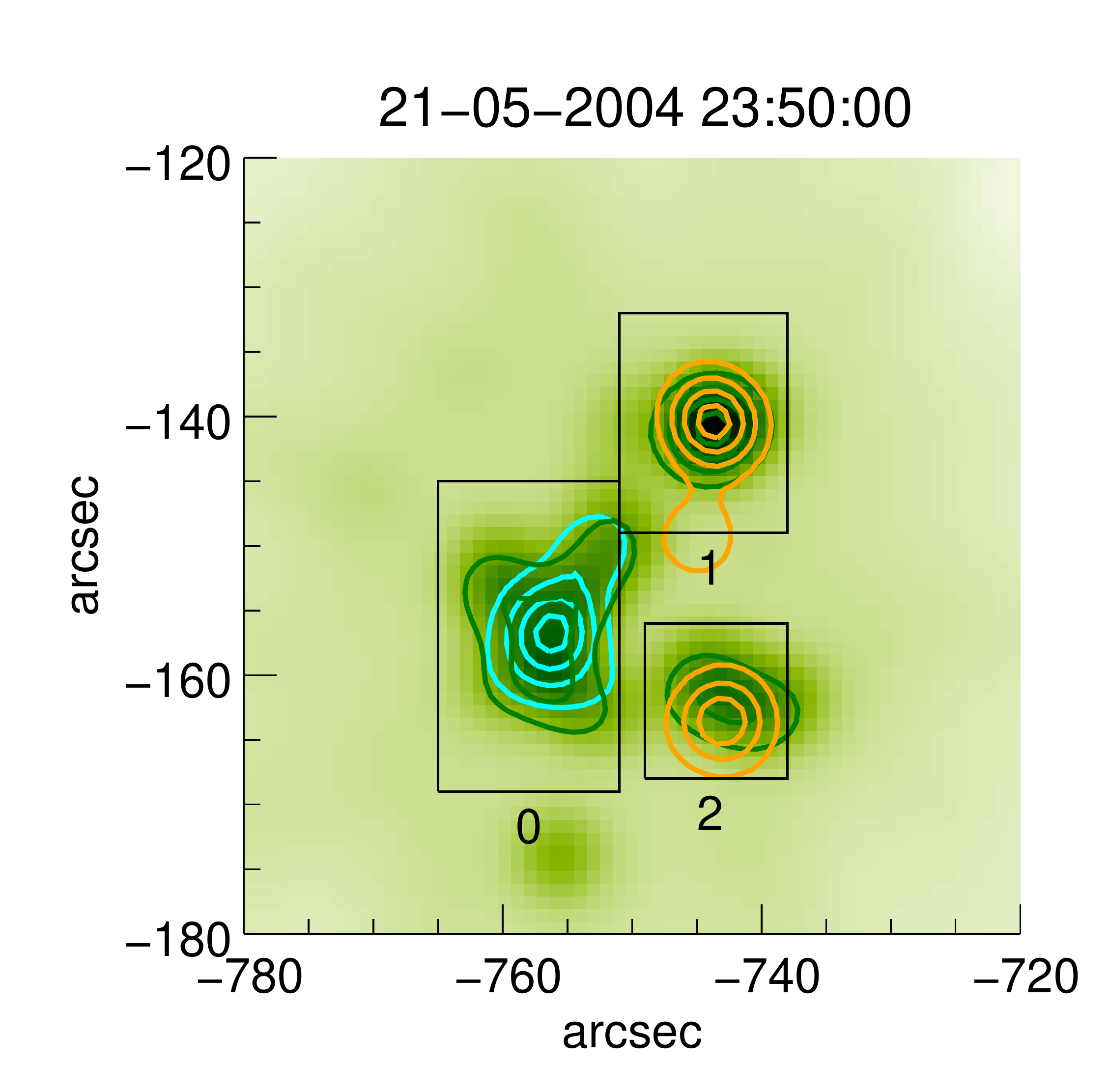}
\caption{CLEAN image (beam factor 1.7) between 23:49:30 and 23:50:30 UT, at 25-50 keV. Contours are 30\%, 50\%, 70\% and 90\% of CLEAN images at 10-25 keV (blue), 25-50 keV (green) and 50-100 keV (orange). Boxes 0, 1, 2 are used for imaging spectroscopy of the looptop source, the first footpoint and the second footpoint respectively. }
\label{clean_images}
\end{center}
\end{figure}

Image and contours at 12-25, 25-50 and 50-100 keV are presented in figure \ref{clean_images}. The geometry of the source can be interpreted as a single loop structure with two footpoints. A coronal hard-X-ray source is visible on the top of the loop structure at 12-25 and 25-50 keV, and the two footpoints are visible in the 25-50 and 50-100 keV ranges.
The loop was divided in three regions (see figure \ref{clean_images}) in order to do imaging spectroscopy on the looptop source and on the two footpoints.
The image reconstruction was done over a 60-second time interval during the main hard X-ray peak, between 23:49:30 and 23:50:30 UT, using the CLEAN algorithm \citep{rhessi_imaging} with a beam factor value of 1.7. The beam factor was carefully chosen as it has an important impact on the determination of the source sizes (see section \ref{section_sizes} and appendix \ref{app_bf}). All collimators except the first one (with the smallest pitch) were used, achieving a spatial resolution of 3.9 arcsec.

\begin{figure*}[ht]
\begin{center}
		\includegraphics[width=0.8\linewidth]{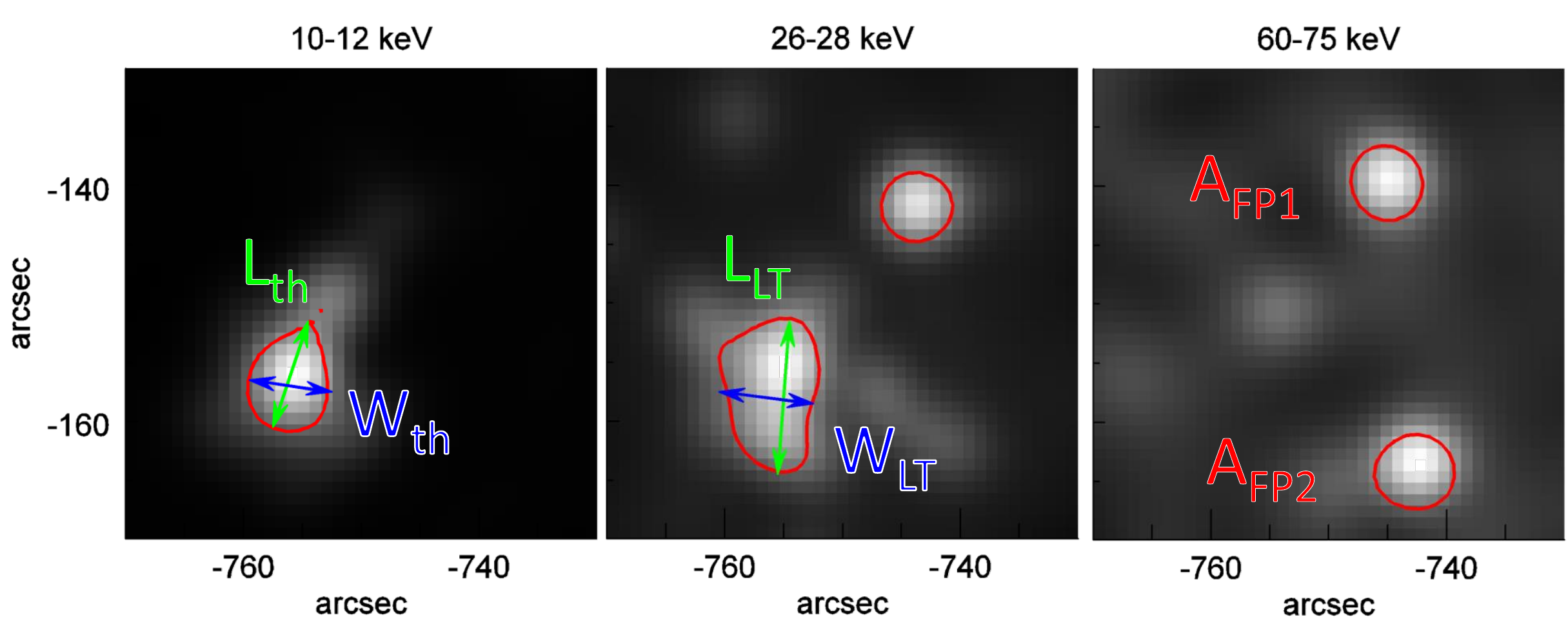}
\caption{RHESSI images in 3 of the 20 energy bins used for imaging spectroscopy, integrated between 23:49:30 and 23:50:30 UT, with the 50 \% of the maximum value enlightened in red. Source length and width are shown with green and blue arrows respectively. At 10-12 keV, the X-ray emission is thermal and we therefore show the length and width of the thermal source $L_{th}$ and $W_{th}$. At 26-28 keV, the emission is non-thermal and we therefore show the lenght and width of the non-thermal looptop source $L_{LT}$ and $W_{LT}$. At 60-75 keV, the area of the footpoint sources is calculated with the 50 \% contour in red.}
	\label{clean_cube}
	\end{center}
\end{figure*}

To do imaging spectroscopy, we reconstructed CLEAN images in 20 narrow energy bins between 10 and 100 keV, with increasing width of the bins with energy (2 keV width between 10 and 30 keV, 3 keV width between 30 and 45 keV, 5 keV width between 45 and 60 keV, 15 keV bin between 60 and 75 keV and 25 keV bin between 75 and 100 keV). Three images over the 20 images produced are presented in figure \ref{clean_cube}, with the 50 \% contours in red. On these images, the looptop source is visible between 10 and 36 keV, and the footpoints are visible above 28 keV. The visibility of looptop and footpoint sources in the images is of course limited by the dynamic range of the images.

\subsection{Spectral analysis}
 
\begin{figure*}
\begin{center}
\includegraphics[width=0.95\linewidth]{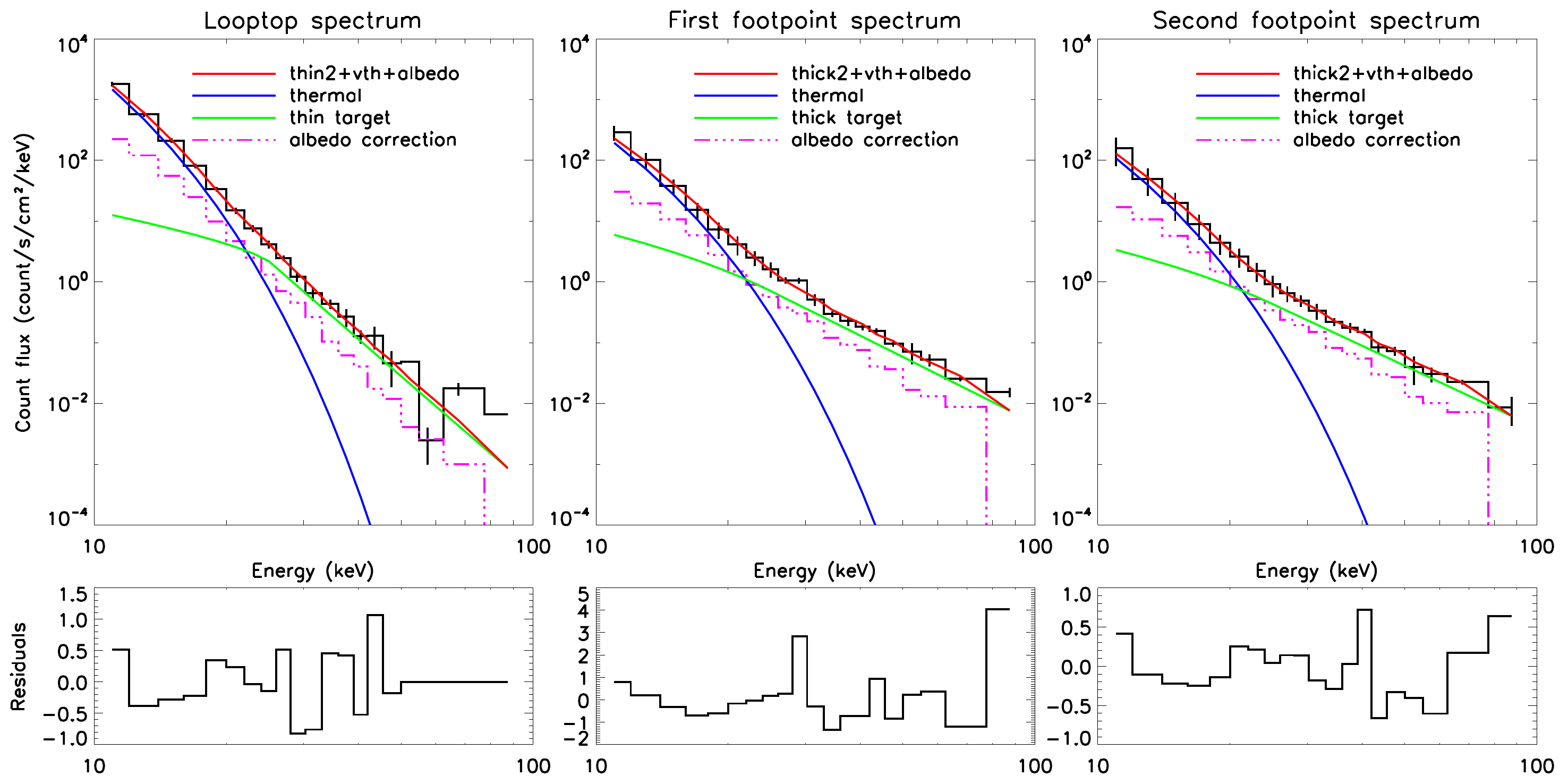}
\caption{Count flux spectra (data and fit) with residuals, for the looptop source (left), the first footpoint (middle) and the second foopoints (right), as defined by the black boxes in figure \ref{clean_images}. The spectra derived from the data is shown in black. The blue curve represent the thermal component of the fit, the green curve represents the non-thermal component. The pink dashed-dotted line represents the component due to albedo correction. The red curve is the total fitted spectrum.}
\label{spectra}
\end{center}
\end{figure*}

Each of the 20 images reconstructed between 10 and 100 keV contributes to a single point in the spectrum. 
The spectra of each region defined in figure \ref{clean_images} were fitted using a combination of a thermal and a non-thermal components in OSPEX \citep{ospex}.
The three spectra resulting from the fits are displayed in figure \ref{spectra} and the values of the free parameters are described in table \ref{spectral_analysis_tab}.

The thermal model has two free parameters which are adjusted during the fit: the temperature and emission measure of the X-ray emitting plasma. 
The non-thermal part of the spectra was fitted with two different models computing the X-ray flux from a single power-law distribution of energetic electrons. 
In the looptop source (region 0 in figure \ref{clean_images}), we assume for simplicity that energetic electrons lose only a small portion of their energy through collisions and that the region can be considered as a thin target. The free parameters of the thin target model are the electronic spectral index $\delta_{LT}$ and a normalisation factor $ \left\langle \bar{n}V\bar{F}_{0}\right\rangle = \left( \int^{\infty}_{E_{0}} \left\langle \bar{n}V\bar{F}(E) \right\rangle dE \right)$ (electrons$/$s$/$cm$^{2}$), where $\bar{n}$ is the mean density of the thin target, $V$ is its volume, and $\bar{F}(E)$ is the energetic electron mean spectrum in electrons$/$s$/$cm$^{2}/$keV (see equation \ref{mean_flux_thin} in appendix).
In the footpoints (regions 1 and 2 in figure \ref{clean_images}), the density is much higher and the energetic electrons lose instantaneously all their energy in the target, considered as a thick target. The free parameters of the thick target model are the electronic spectral index $\delta_{FP}$ and the electron rate above $E_0$, $\dot{N}$ (electrons$/$s), entering the target (see equation \ref{mean_flux_thick} in appendix).
In each case, a minimum correction for albedo was used (assuming an isotropic beam of electrons), and the low energy cutoff $E_{0}$ of the non-thermal model (thick or thin target models) was fixed to 25 keV, since when this parameter was set free in the spectral analysis, it reached 23 keV.

\setlength{\tabcolsep}{0.1cm} % cette commande permet de choisir l'espacement entre les colonnes d'un tableau
\renewcommand{\arraystretch}{1.3} % et la c'est pour l'espacement entre lignes

\begin{table}
\caption[10pt]{Values of the free spectral parameters obtained for the looptop source, and for the first and second footpoints. EM and T are the emission measure and the temperature (thermal component). $\left\langle \bar{n}V\bar{F}_{0} \right\rangle$ is the normalisation factor derived for the looptop source in the thin target approximation, $\dot{N}$ is the electron rate above 25 keV derived for the footpoints in the thick target approximation, $\delta$ is the electron spectral index derived in both thin and thick target approximations. Note that the non-thermal parameters refers to the electron distribution directly and not to the photon spectrum.}
\label{spectral_analysis_tab}
\begin{center}
\begin{tabular}{@{}cccc@{}}
\hline
                                &                 & First           & Second          \\
																& Looptop         & Footpoint       & Footpoint       \\
\hline
\hline
EM ($\times 10^{48}$ cm$^{-3}$) & $2.1 \pm 0.5$ & $0.14 \pm 0.08$ & $0.08 \pm 0.07$ \\
\hline
T (keV)                         & $2.1 \pm 0.1$   & $2.5 \pm 0.3$   & $2.5 \pm 0.4$   \\
T ($\times 10^{6}$ K)           & $24 \pm 1.2$    & $29 \pm 3.5$    & $29 \pm 4.6$   \\
\hline
\hline
$\left\langle \bar{n}V\bar{F}_{0}\right\rangle > 25$ keV        &                 &                 &                 \\
($\times 10^{55}$ e$^{-}$ cm$^{-2}$ s$^{-1}$) & $0.46 \pm 0.08$ &                 &                 \\
\hline
$\dot{N}$ ($\times 10^{35}$ e$^{-}$ s$^{-1}$) &                 & $0.12 \pm 0.03$ & $0.06 \pm 0.02$ \\
\hline
$\delta$                                      & $5.2 \pm 0.4$   & $4.4 \pm 0.2$   & $4.2 \pm 0.2$   \\
\hline
\hline
\end{tabular}
\end{center}
\end{table}

\subsection{Sizes and density of the thermal and non-thermal X-ray sources}
\label{section_sizes}

In the further calculation of the electron rate for the different X-ray sources (see section \ref{section_electron_rate}), we need to estimate the sizes of the different X-ray emitting regions: the coronal source and the footpoints. Moreover, we distinguish the thermal X-ray emitting region from the non-thermal X-ray emitting region in the coronal source. 
We use the 50\% CLEAN contours from the images to estimate the length, width or area of the X-ray sources. The CLEAN images were produced with a beam factor of 1.7. The determination and the influence of this parameter are discussed in appendix \ref{app_bf}. 

\begin{figure}[t]
\begin{center}
		\includegraphics[width=0.86\linewidth]{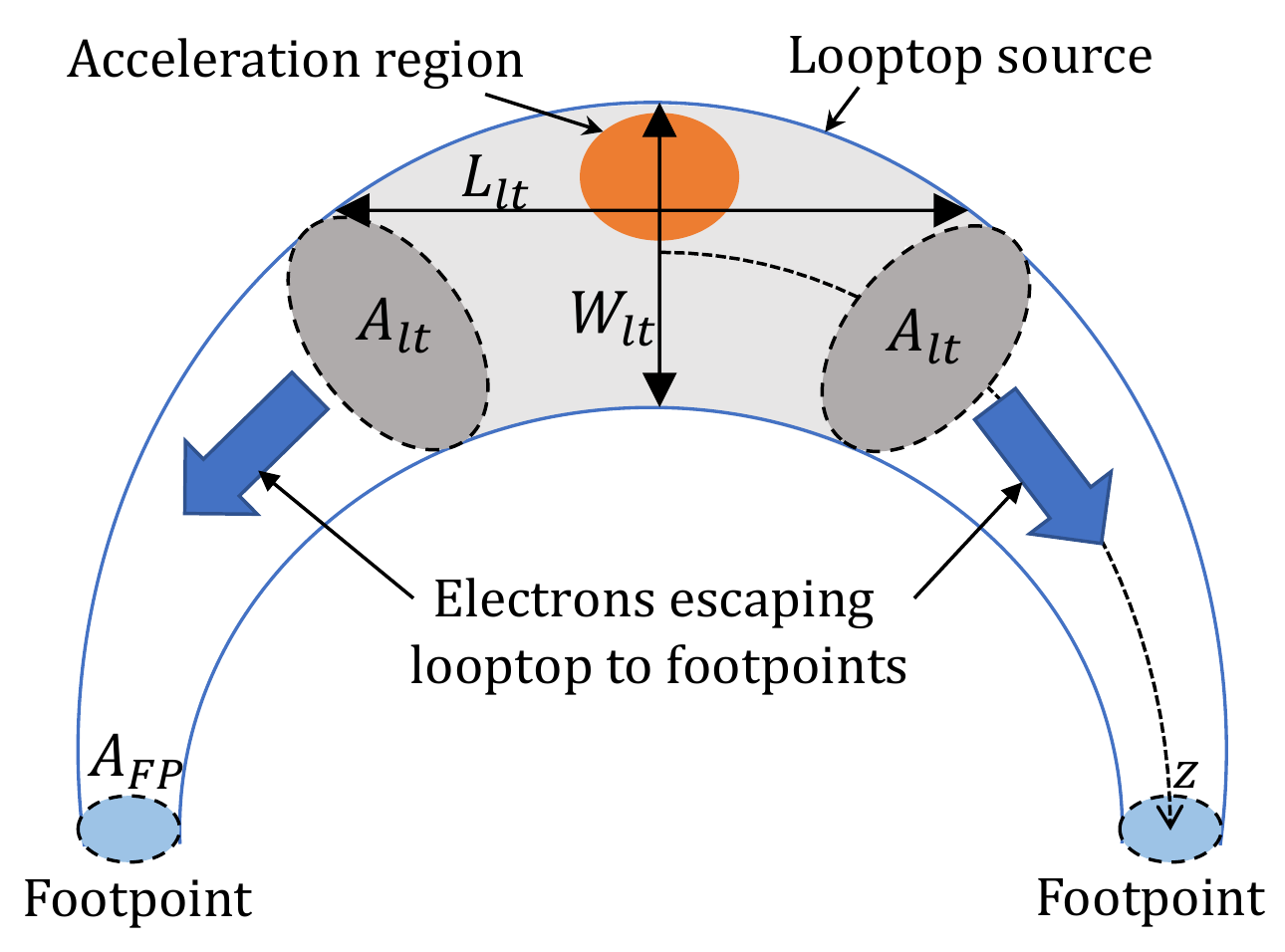}
	\caption{Sketch of a symmetrical magnetic loop. The limits of the looptop sources are the cross-sections of the loop with area A$_{LT}$ shown in grey. The length L$_{LT}$ and the width W$_{LT}$ of the looptop X-ray source are used to determine the size of the looptop source, which is approximated to a cylinder of diameter W$_{LT}$. Blue arrows represent the electron rate for electrons leaving the looptop source of cross-section A$_{LT}$.}
	\label{schema_loop}
	\end{center}
\end{figure}

The size of the thermal coronal source was measured at 10-12 keV, to ensure the X-ray emission to be entirely thermal (see the looptop spectrum, on the left panel in figure \ref{spectra}). The size of the non-thermal X-ray source at the looptop is measured at 26-28 keV, since at this energy, the looptop source is still visible in the image, and the X-ray spectra is predominantly non-thermal (see the looptop spectrum, on the left panel in figure \ref{spectra}). Finally, the area of the footpoints is taken at 60-75 keV. The measurements of the length and width of the coronal sources are represented by green and blue arrows respectively in figure \ref{clean_cube} and schematically explained in figure \ref{schema_loop}. The measured sizes and areas are summarized in table \ref{size_tab}. 

\begin{table}
\caption[10pt]{Measured sizes of the thermal and non-thermal (looptop and footpoints) sources, using the 50\% contours of the CLEAN images as shown in figure \ref{clean_cube}.}
\label{size_tab}
\begin{center}
\begin{tabular}{@{}cccc@{}}
\hline
 & Width & Length & Area \\
 & (Mm)  & (Mm)   & (Mm$^{2}$) \\
\hline
Thermal source (10-12 keV)  & 5.3 & 7.0 & \\
Looptoop source (26-28 keV) & 5.8 & 9.6 &  \\
1st Footpoint (60-75 keV)  & & & 17.6 \\
2nd Footpoint (60-75 keV) & & & 19.2 \\
\hline
\end{tabular}
\end{center}
\end{table}

The emission measure $EM$ given by the spectral analysis (see table \ref{spectral_analysis_tab}) of the thermal part of the coronal source, and the estimation of the size of the thermal source (see table \ref{size_tab}), leads to the following estimation of the density:
\begin{equation}
\bar{n} = \sqrt{\frac{EM}{V_{th}}} = \sqrt{\frac{EM}{L_{th}A_{th}}} 
\end{equation}
where $\bar{n}$ is the density (in cm$^{-3}$), $V_{th}$ is the volume of the thermal source (in cm$^{3}$). $L_{th}$, $W_{th}$ and $A_{th}$ are respectively the length, the width and the cross-section of the thermal source, with $A_{th} = \pi \left( W_{th}/2 \right)^{2}$. The assumed geometry of the loop is described in figure \ref{schema_loop}.

The mean plasma density obtained is $\bar{n} = \left( 1.2 \pm 0.2 \right) \times 10^{11}$ cm$^{-3}$. Note that this value is in the range of densities calculated by \cite{simoes_and_kontar_2013} for events where a non-thermal looptop source is visible.

%%%%%%%%%%%%%%%%%%%%%%%%%%%%%%%%
% *** DATA AND METHODOLOGY *** %
%%%%%%%%%%%%%%%%%%%%%%%%%%%%%%%%

\section{Determination of the spatial and spectral distributions of X-ray emitting energetic electrons}
\label{section3}

\subsection{Comparison of electron rates}
\label{section_electron_rate}

As explained below, the electron rate above $E_0 = 25$ keV of electrons leaving the looptop source is found to be about $\dot{N}_{LT} = \left( 0.4 \pm 0.2 \right) \times 10^{35}$ electrons s$^{-1}$.

Indeed, the electron rate $\dot{N}_{LT}$ (in electrons s$^{-1}$) in the looptop source is given by: 
\begin{equation}
\dot{N}_{LT} = A_{LT} \int^{\infty}_{E_{0}} \bar{F}(E) dE  
\end{equation}
Where $E_{0}$ is the low energy cutoff (in keV), $A_{LT}$ (cm$^{2}$) is the cross-section of the looptop source (assuming a symmetrical source) as shown in figure \ref{schema_loop}, and $\bar{F}(E)$ is the mean energetic electron spectrum (in electrons s$^{-1}$ cm$^{-2}$ keV$^{-1}$). We assume energetic electrons propagating in both directions along the loop axis (see the blue arrows in figure \ref{schema_loop}).

For a source with an homogeneous ambiant plasma, we can express the looptop electron rate as follows:
\begin{equation}
\dot{N}_{LT} = A_{LT} \int^{\infty}_{E_{0}} \frac{\left\langle  \bar{n}V\bar{F} \right\rangle}{\bar{n}V} dE = \frac{1}{\bar{n} L_{LT}} \int^{\infty}_{E_{0}} \left\langle \bar{n}V\bar{F} \right\rangle dE
\end{equation}
$\left\langle \bar{n}V\bar{F}_{0} \right\rangle = \int^{\infty}_{E_{0}} \left\langle  \bar{n}V\bar{F} \right\rangle dE$ is given by the spectral analysis of the looptop source (see table \ref{spectral_analysis_tab}) and $L_{LT}$ has been measured on the 26-28 keV CLEAN image (see table \ref{size_tab}). 

The electron rate obtained for the looptop source, $\dot{N}_{LT} = \left( 0.4 \pm 0.2 \right) \times 10^{35}$ electrons s$^{-1}$, is compared to the electrons rates obtained by the spectral analysis of the two footpoints, which are $\left( 0.12 \pm 0.03 \right) \times 10^{35}$ and $\left( 0.06 \pm 0.02 \right) \times 10^{35}$ electrons s$^{-1}$. 
The sum of the rates from the footpoints is therefore significantly lower than the rate needed to explain the nonthermal emission in the coronal source: the ratio $\frac{\dot{N}_{LT}}{\dot{N}_{FP}}$ is about 2.2 for this event, for electrons above $E_0=25$ keV.

%%%%%%%%%%%%%%%%%%%%%%%%%%%%%%%%%%%%%%%%%%%%%%%%%%%%%%%%%%%%%%%%%%%%%%%%%%%%%%%%%%%%%%%%%%%%%%%%%%%%%%%%%%%%%%%%%%%%%%%%%%

\subsection{Spatial and spectral distributions of the mean flux spectrum and the density of energetic electrons}
\label{section_spatial_distrib_and_spectra}

\begin{figure}
\begin{center}
\includegraphics[width=0.99\linewidth]{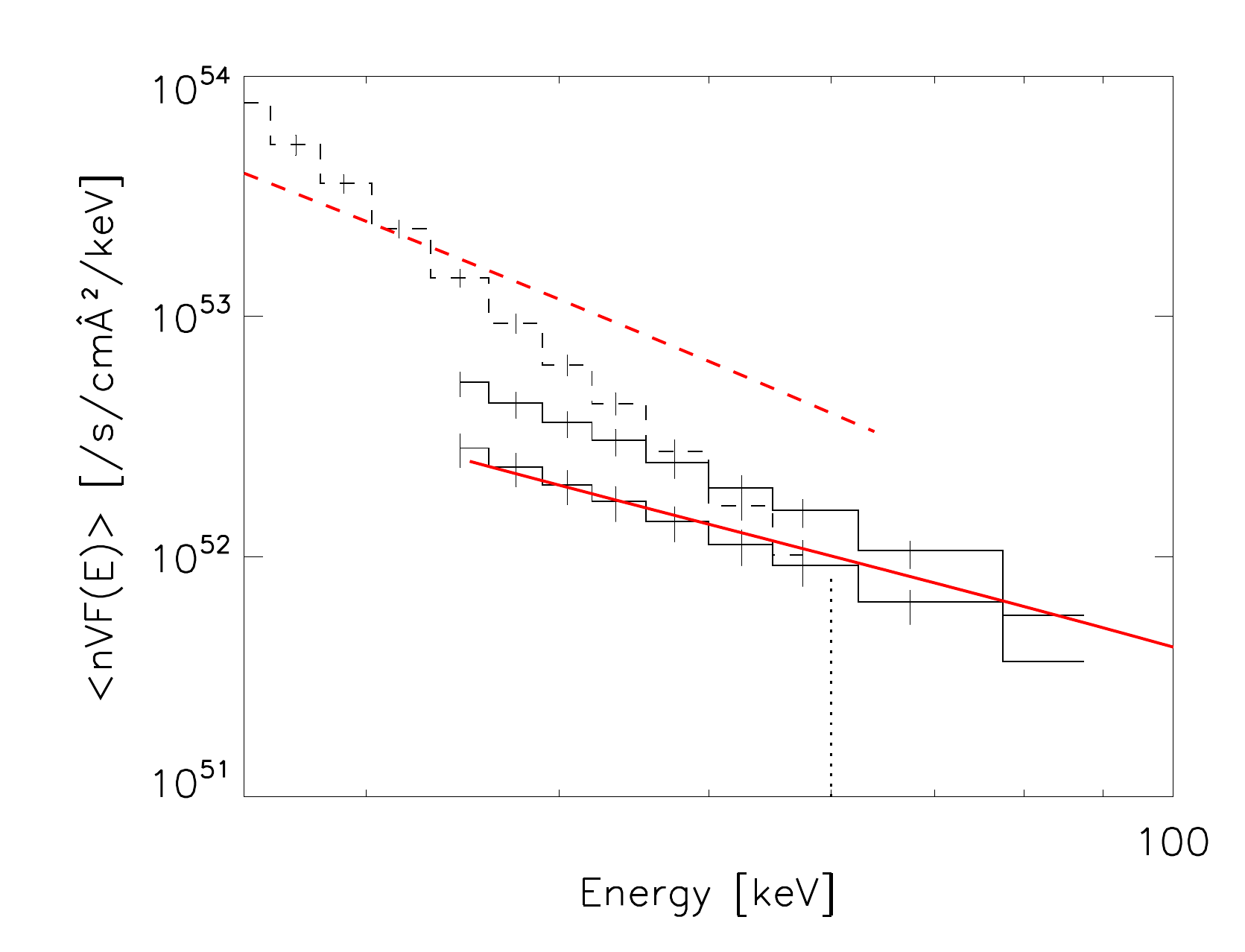}
\caption{Spatially integrated and density weighted mean flux spectra for the looptop (dashed line) and the footpoint sources (plain lines) deduced from X-ray observations (black histograms), and computed with the diffusive transport of \cite{kontar_et_al_2014} model with $n = 9.5 \times 10^{10}$ cm$^{-3}$, $d = 5.5$ Mm and $\lambda = 1.4 \times 10^8$ cm (red). The dotted vertical line marks the energy at which the coronal and the second footpoint spectra cross.}
\label{nvF}
\end{center}
\end{figure}

Using equations \ref{mean_flux_thin} and \ref{mean_flux_thick} (in appendix \ref{app_eqn}) and the results of the spectral analysis displayed in table \ref{spectral_analysis_tab}, the spatially integrated density weighted mean flux spectra $\left\langle nVF(E)\right\rangle$
of the looptop source and footpoints are plotted on figure \ref{nvF}, in black. On this figure, the energies for which the footpoint spectra are crossing the looptop spectrum are 50 keV and 60 keV for the first and the second footpoints respectively.

\begin{figure}
\begin{center}
\includegraphics[width=0.99\linewidth]{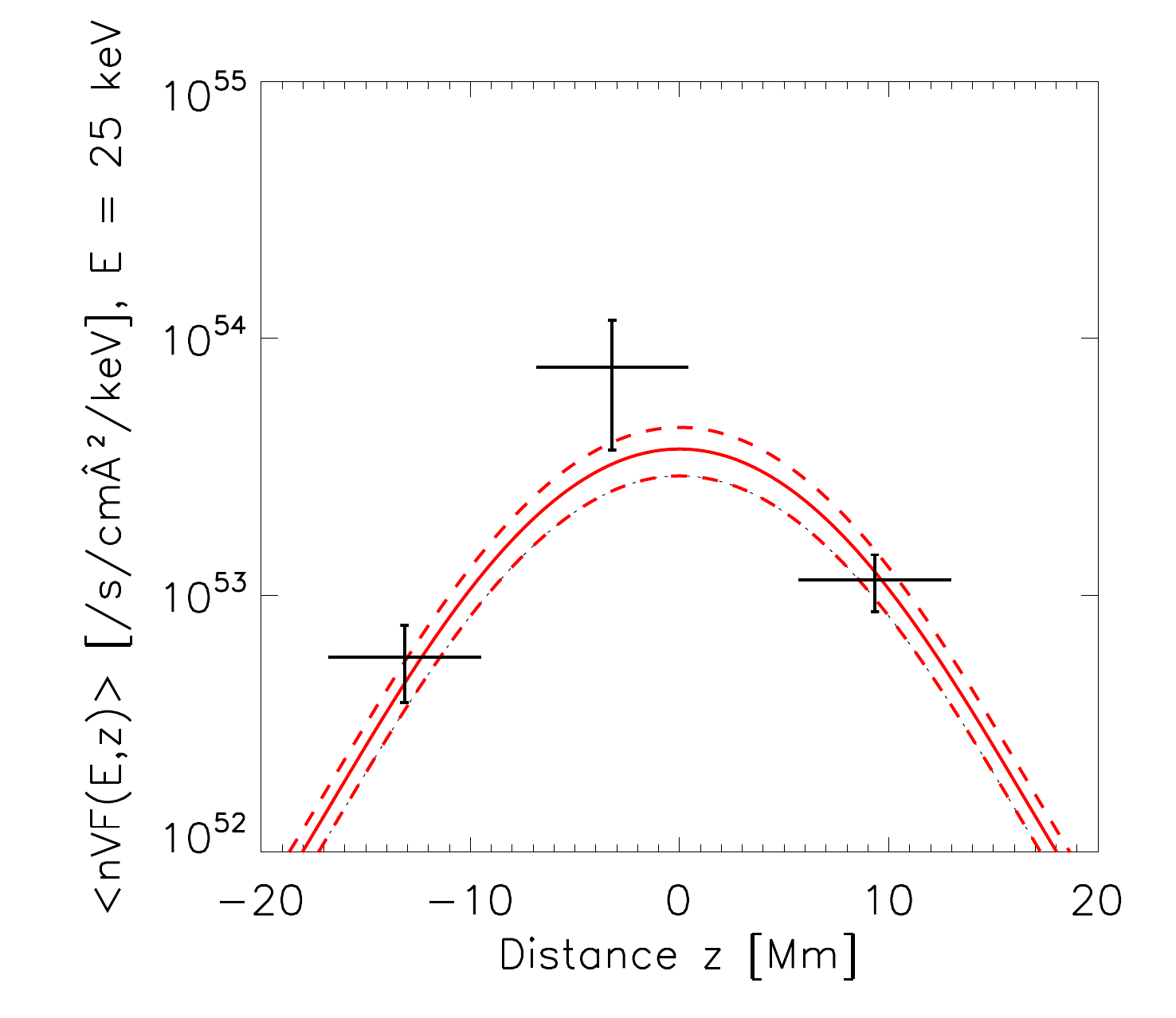}
\caption{Spatial distribution of energetic electrons at 25 keV deduced from X-ray observations (black crosses) and computed with the diffusive transport model of \cite{kontar_et_al_2014} with $n = 9.5 \times 10^{10}$ cm$^{-3}$, $d = 5.5$ Mm and $\lambda = 1.4 \times 10^8$ cm (red lines). The looptop source is shifted of 3.2 Mm in regards to the top of the loop, as it has been described in \cite{kuznetsov_and_kontar_2015}. For the model, the dashed or dotted lines mark a confidence interval around the computed value marked by the plain line. The detailed description is in section \ref{interpretation_and_model_prediction}.}
\label{spatial_nvf}
\end{center}
\end{figure}

The spatial distribution of energetic electrons $\left\langle nVF \right\rangle$ at 25 keV is also known in three locations in the loop (looptop and footpoints). The distance between the footpoints and the looptop source is estimated by taking the distance between the centers of the boxes defined in figure \ref{clean_images}. \cite{kuznetsov_and_kontar_2015} showed that the maximum of the spatial distribution of energetic electrons was shifted of 3.2 Mm in regards with the top of the magnetic loop. This three-point distribution is shown in figure \ref{spatial_nvf}.

%%%%%%%%%%%%%%%%%%%%%%%%%%%%%%%%%%%%%%%%%%%%%%%%%%%%%%%%%%%%%%%%%%%%%%%%%%%%%%%%%%%%%%%%%%%%%%%%%%%%%%%%%%%%%%%%%%%%%%%%%%

\label{section_spatial_distribution_density}

The number density of energetic electrons with energy $E > E_{min}$, $n_{b}^{E_{min}}$ (in electrons cm$^{-3}$), is defined as:
\begin{equation}
n_{b}^{E_{min}} \equiv \int^{\infty}_{E_{min}} \frac{F(E)}{v(E)} dE
\label{density_def}
\end{equation}
where $v$ is the velocity of the electrons.
In the following, we distinguish the estimation of $n_{b}^{E_{min}}$ in the thin and in the thick target models. The details of the calculations are in appendix \ref{app_eqn}.

\begin{figure}
\begin{center}
\includegraphics[width=0.99\linewidth]{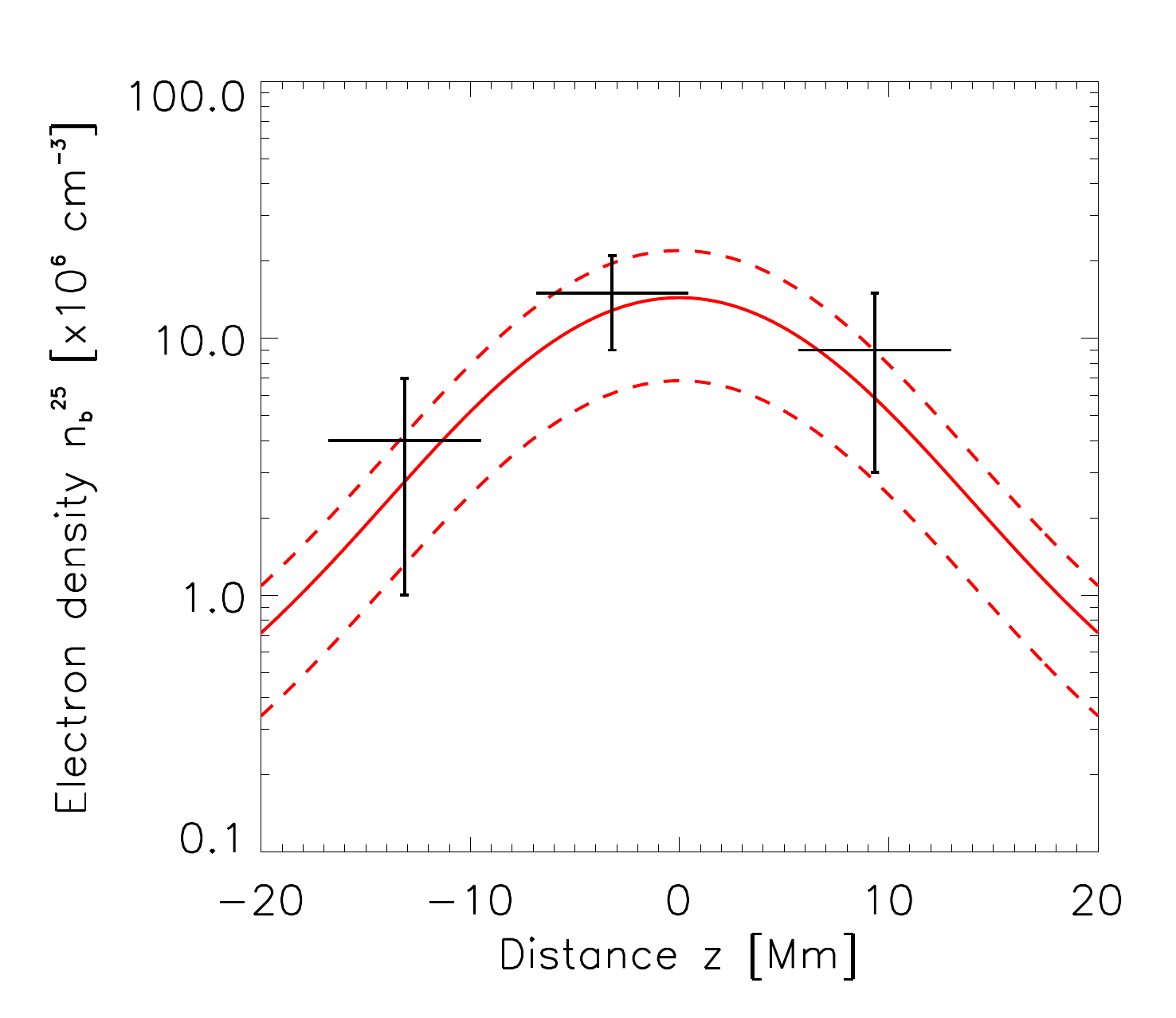}
\caption{Spatial distribution of the density of energetic electrons with energy $E > 25$ keV, deduced from observations ($n_b^{25}$, black crosses). The spatial distributions of $n_b^{25}$ calculated with the diffusive transport model of \cite{kontar_et_al_2014}, with $A_{LT} = 26$ Mm$^2$, $d = 5.5$ Mm, $n = 9.5 \times 10^{10}$ cm$^{-3}$ and $\lambda = 1.4 \times 10^8$ cm (red lines) is also plotted. The looptop source is shifted of 3.2 Mm in regards to the top of the loop, as it has been described in \cite{kuznetsov_and_kontar_2015}. The detailed description is in section \ref{interpretation_and_model_prediction}.}
\label{spatial_distribution_25}
\end{center}
\end{figure}

Using equations \ref{nb_thin} and \ref{nb_thick} with $E_{min} = E_{0} = 25$ keV, we can evaluate the electron density of energetic electrons with energy $E > 25$ keV, $n_{b}^{25}$ in the thin and thick target models respectively.
We found $n_{b}^{25} = (15 \pm 6) \times 10^{6}$ electrons cm$^{-3}$ in the corona and $n_{b}^{25} = (9 \pm 6) \times 10^{6}$ and $(4 \pm 3) \times 10^{6}$ electrons cm$^{-3}$ in each footpoint. 
Note that to calculate the number density of energetic electrons from the observations, we need an estimation of the area of the cross-section of the loop $A_{LT}$. From the size estimation displayed in table \ref{size_tab}, we found $A_{LT} = \pi \left( W_{LT}/2 \right)^{2} = 26$ Mm$^2$.
The spatial distribution of the energetic electron density above 25 keV electrons in the flaring loop is plotted on figure \ref{spatial_distribution_25}.

\begin{figure}
\begin{center}
\includegraphics[width=0.99\linewidth]{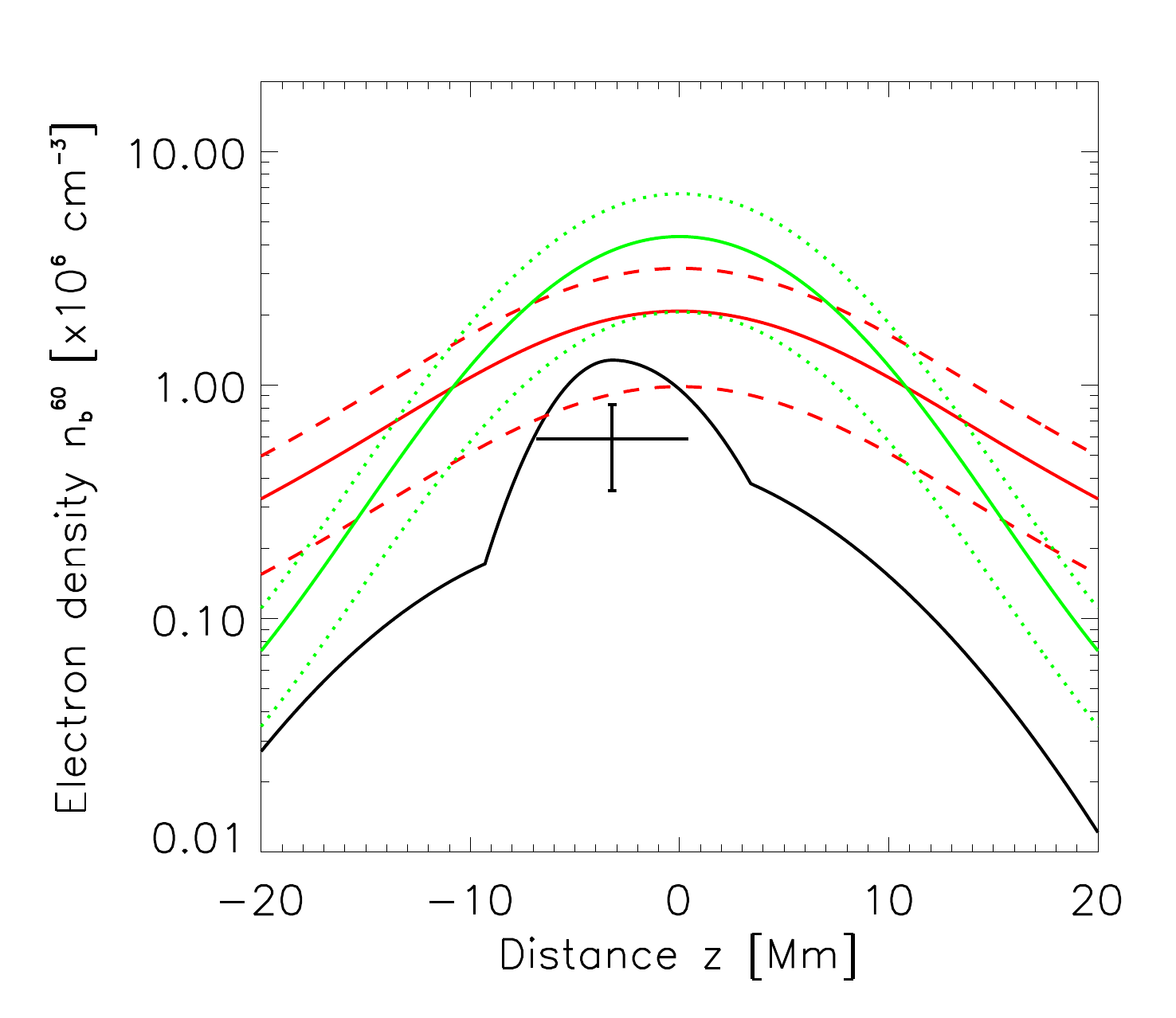}
\caption{Spatial distribution of the density of energetic electrons with energy $E > 60$ keV, at looptop, estimated from X-ray observations ($n_b^{60}$, black cross) and radio observations \citep[black plain line, see][]{kuznetsov_and_kontar_2015}. The spatial distributions of $n_b^{60}$ calculated with the diffusive transport model of \cite{kontar_et_al_2014}, with $A_{LT} = 26$ Mm$^2$, $d = 5.5$ Mm, $n = 9.5 \times 10^{10}$ cm$^{-3}$ and $\lambda = 1.4 \times 10^8$ cm (red lines) and $\lambda = 10^7$ cm (green lines), are also plotted. The looptop source is shifted of 3.2 Mm in regards to the top of the loop, as it has been described in \cite{kuznetsov_and_kontar_2015}. The detailed description is in section \ref{interpretation_and_model_prediction}.}
\label{spatial_distribution_60}
\end{center}
\end{figure}

%%%%%%%%%%%%%%%%%%%%%%%%%%%%%%%%%%%%%%%
% *** DISCUSSION - INTERPRETATION *** %
%%%%%%%%%%%%%%%%%%%%%%%%%%%%%%%%%%%%%%%

\section{Interpretation of the observations in the context of diffusive transport of energetic electrons}
\label{interpretation}

\subsection{Confinement of X-ray producing energetic electrons}
\label{interpretation_and_model_prediction}
%----------

The spectral indexes of the electron distribution from the X-ray emission, using the thin and thick target models are summarized in table \ref{spectral_analysis_tab}. While electron spectral indexes in both footpoints are very close and will be considered as similar, the electron spectral index in the loop top source is 
softer by $\approx 1$. Furthermore the ratio $\frac{\dot{N}_{LT}}{\dot{N}_{FP}}$ of the electron rate in the looptop source and in the footpoints is found to be around 2.2. These values are similar to values found for other events \citep[see e.g.][]{simoes_and_kontar_2013}. These results suggest that a significant number of high energy electrons are confined in the coronal region. Such a confinement of high energy electrons can result from magnetic mirroring or turbulent pitch angle scattering as  demonstrated in \cite{kontar_et_al_2014, bian_et_al_2011}. In the following we will investigate whether the confinement observed in this flare can be explained in the context of the diffusive transport model of \cite{kontar_et_al_2014}. In this model, strong turbulent pitch angle scattering, due to small scale fluctuations of the magnetic field, is responsible for a diffusive parallel transport of energetic particules and finally results as a confinement mechanism. 

\subsubsection{Effects of the diffusive transport on energetic electron distributions}
\label{discussion_model}

In section \ref{model_fitting} we will compare the observed spatial and spectral distributions of $\left\langle nVF\right\rangle$ with the ones calculated in the diffusive transport model described in \cite{kontar_et_al_2014}. The distribution $F_D(E,z)$ (electrons$/$cm$^2/$s$/$keV) of energetic electrons of energy $E$ at a position $z$ along the magnetic loop is indeed described by the following equation \citep{kontar_et_al_2014}:
\begin{equation}
\begin{split}
F_D(E,z) = \frac{E}{Kn}\int_E^{\infty} dE' \frac{F_0(E')}{\sqrt{4a\pi(E'^2-E^2)+2d^2}} \\
\times \exp{\left(\frac{-z^2}{4a(E'^2-E^2)+ 2d^2}\right)}
\end{split}
\label{equation}
\end{equation}
Where $F_0(E)$ is the initial distribution of energetic electrons in the source (acceleration) region supposed to be spatially extended, $d$ is the size of the acceleration region (Gaussian form), $n$ is the plasma ambiant density, $K=2\pi e^4 \Lambda$ is the collisional parameter, and $a=\lambda/(6Kn)$ where $\lambda$ is the pitch-angle scattering mean free path of the electrons. In the work of \cite{kontar_et_al_2014}, this mean free path $\lambda$ is considered to be independent of energy.

\begin{figure*}
\begin{center}
\includegraphics[width=0.86\linewidth]{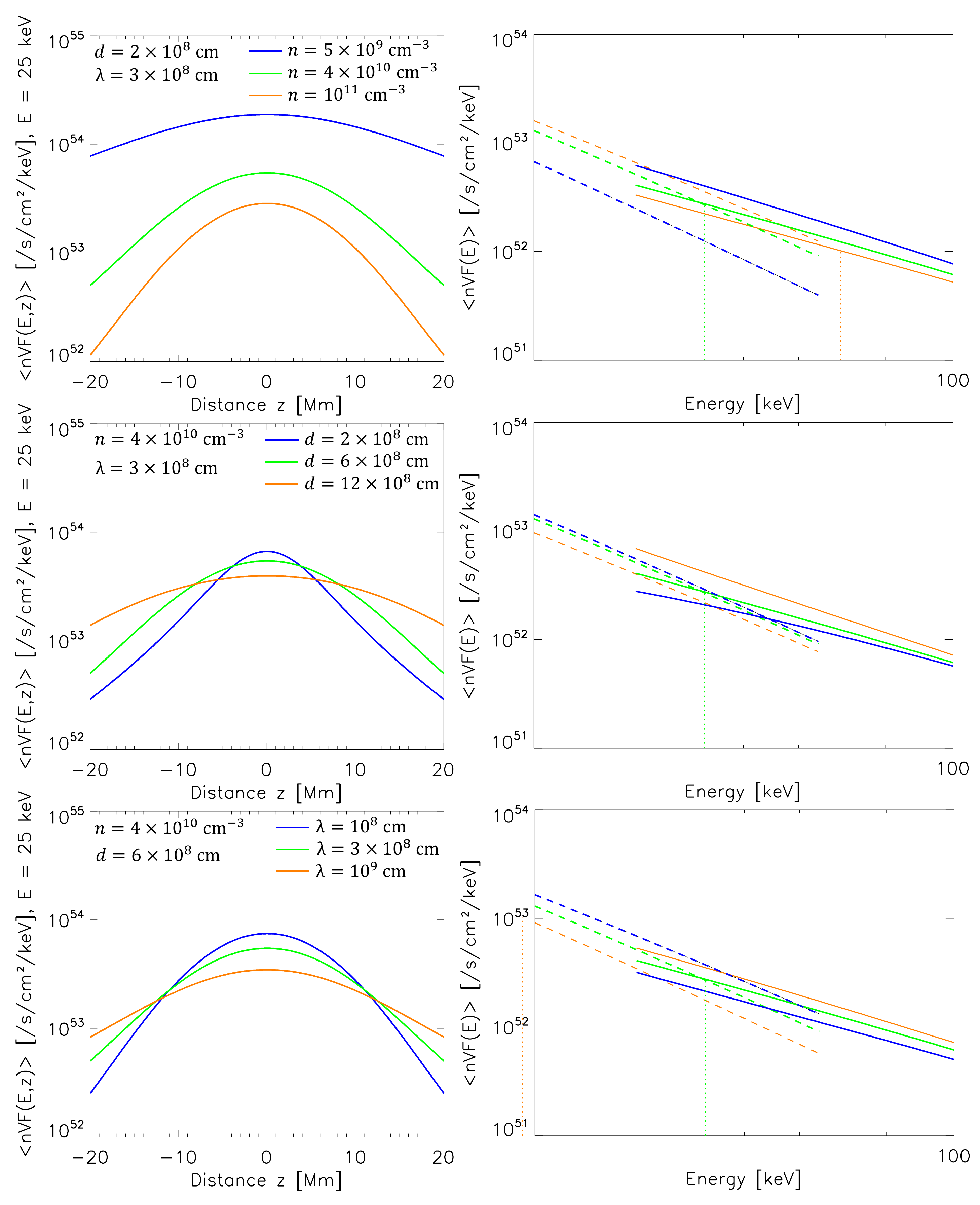}
\caption{Influence of the free parameters in equation \ref{equation} ($n$ the plasma density, $d$ the size of the acceleration region, and $\lambda$ the scattering mean free path) on the spatial (left panels) and spectral (right panels) distributions of energetic electrons. In the top panels, $d$ and $\lambda$ are constant and $n = 5 \times 10^{9}$ cm$^{-3}$ (blue), $n = 4 \times 10^{10}$ cm$^{-3}$ (green) and $n = 10^{11}$ cm$^{-3}$ (orange). In the middle panels, $n$ and $\lambda$ are constant and $d=2 \times 10^8$ cm (blue), $d=6 \times 10^8$ cm (green) and $d= 12 \times 10^8$ cm (orange). In the bottom panels, $n$ and $d$ are constant and $\lambda= 10^8$ cm (blue), $\lambda= 3 \times 10^8$ cm (green) and $\lambda= 10^9$ cm (orange). The dotted vertical lines mark the energies at which the coronal spectrum crosses the footpoint spectrum.}
\label{influence_param}
\end{center}
\end{figure*}

In equation \ref{equation} the plasma density $n$, the size of the acceleration region $d$ and the electron scattering mean free path $\lambda$ are parameters to the model. The effects of these parameters on both spatial and spectral distributions of energetic electrons is shown in figure \ref{influence_param} and summarized in table \ref{table_param}. The spectral indexes obtained in the corona and in the footpoints for each set of parameters are described in table \ref{table_indexes}.

\begin{table*}
\caption[10pt]{Summary of the influence of the density $n$, the size of the acceleration region $d$ and the scattering mean free path $\lambda$ on the spatial and spectral distributions of energetic electrons in the frame of the diffusive transport model.}
\label{table_param}
\begin{center}
\begin{tabular}{@{}ccc@{}}
\hline
 Parameter                & Effect on spatial distribution & Effect on spectra \\
\hline
 When $n$ increases       & The spatial distribution gets narrower and & The spectra gets harder and the   \\
                          & the peak of the distribution decreases     & energy at which coronal and footpoint  \\
					                &                                            & spectra cross increases \\
 When $d$ increases       & The spatial distribution gets broader and  & The footpoint spectrum gets softer and  \\
                          & the peak of the distribution decreases     & the energy at which coronal and footpoint \\
                          &                                            & spectra cross decreases \\
 When $\lambda$ increases & The spatial distribution gets broader and  & The spectra get softer and  \\
                          & the peak of the distribution decreases     & the energy at which coronal and foortpoint\\ 
							          	&                                            & spectra cross decreases \\
\hline
\end{tabular}
\end{center}
\end{table*}

\begin{table}
\caption[10pt]{Influence of the density $n$, the size of the acceleration region $d$ and the scattering mean free path $\lambda$ on the spectral index of non-thermal electron distributions in the frame of the diffusive transport model. The spectral indexes $\delta_C$ of the coronal spectra is measured with a linear regression between 25 and 65 keV; the spectral indexes $\delta_{FP}$ are measured in the same way between 50 keV and 100 keV. When $n$ varies, $d=2 \times 10^8$ cm and $\lambda=3 \times 10^8$ cm; when $d$ varies, $n=4 \times 10^{10}$ cm$^{-3}$ and $\lambda=3 \times 10^8$ cm; and when $\lambda$ varies, $n=4 \times 10^{10}$ cm$^{-3}$ and $d=6 \times 10^8$ cm. }
\label{table_indexes}
\begin{center}
\begin{tabular}{@{}cccc@{}}
\hline
 Density $n$                  & $5 \times 10^{9}$           & $4 \times 10^{10}$  & $10^{11}$ \\
\hline
 $\delta_C$                   & $3.02 \pm 0.01$     & $2.84 \pm 0.01$     & $2.72 \pm 0.01$   \\
 $\delta_{FP}$, $E > 50$ keV  & $2.02 \pm 0.01$     & $1.84 \pm 0.01$     & $1.77 \pm 0.01$   \\
\hline
\hline
 Size $d$ (cm)                & $2 \times 10^{8}$  & $6 \times 10^{8}$   & $12 \times 10^{8}$ \\
\hline
 $\delta_C$                   & $2.87 \pm 0.01$     & $2.84 \pm 0.01$     & $2.69 \pm 0.01$   \\
 $\delta_{FP}$, $E > 50$ keV  & $1.63 \pm 0.01$     & $1.84 \pm 0.01$     & $2.14 \pm 0.01$   \\
\hline
\hline
 Lambda $\lambda$ (cm)        & $10^{8}$            & $3 \times 10^{8}$   & $10^{9}$          \\
\hline
 $\delta_C$                   & $2.70 \pm 0.01$     & $2.84 \pm 0.01$     & $2.96 \pm 0.01$   \\
 $\delta_{FP}$, $E > 50$ keV  & $1.76 \pm 0.01$     & $1.84 \pm 0.01$     & $1.95 \pm 0.01$   \\
\hline
\end{tabular}
\end{center}
\end{table}

As shown on figure \ref{influence_param}, the different parameters do not influence the two distributions in the same way. For instance, when $d$ or $\lambda$ increases, the spatial distribution becomes broader (note that the shape of the distribution remains different), but the effects on the spectra are not the same: increasing $d$ will have almost no impact on the coronal spectrum whereas increasing $\lambda$ leads to a softening of the coronal spectrum.

It can be seen that the increase of density leads to an enhanced trapping of energetic electrons, even if the scattering mean free path remains unchanged.

In addition, we note that there is a limit value of the size of the acceleration region below which the effect of this parameter is negligible: this is the case when $d^2 \ll a \pi (E'^2-E^2)$ (see equation \ref{equation}). In our conditions, the influence of $d$ on the energetic electron distributions is negligible for $d \lesssim 10^8$ cm.

\subsubsection{Model fitting to the observed energetic electron distributions}
\label{model_fitting}

In the following, the electron distribution in the loop top and in the footpoints are computed through integration of equation 5 on $z$ (respectively from -7 Mm to 5 Mm for loop top sources and from -9 Mm to -$\infty$ and from +13 Mm to + $\infty$ for footpoint sources). In order to fit the distributions derived from these equations to the observations, we need to determine the injected distribution of electrons $F_0$, and make the parameters $n$, $d$ and $\lambda$ vary. As discussed in \cite{kontar_et_al_2014} and also shown in section \ref{discussion_model}, the increase of diffusion due to pitch-angle scattering results in enhanced coronal emission and weaker footpoint emission than in the standard non diffusive case, due to the increase of the time spent by the electrons in the corona. In the diffusive case, the spectrum of the electrons in the corona becomes progressively flatter, with decreasing scattering mean free path. However, as shown in section \ref{discussion_model}, the mean electron spectrum in the footpoint is less affected by the increase of diffusion that the electron spectrum in the corona. This is why in the following of the paper, we assume that the injection spectrum of the energetic electrons is given by the spectral index of the population of the electrons entering the footpoints. 
The injected electron rate cannot be directly inferred from the results of the spectral analysis of the observations: indeed, the value found in the footpoints is too small because there are trapped electrons while the electron rate computed in the corona (see section \ref{section_electron_rate}) is too large since there is some trapping effects. Therefore, the injected electron rate is considered as a free parameter to fit the model to the observations, with however the constraint that its value $\dot{N}_i$ must be between the two boundaries $\dot{N}_{FP}$ and $\dot{N}_{LT}$. This parameter has no effect on the spatial distribution of energetic electrons but impacts the normalisation of the spatial distribution of the density of energetic electrons. 
 
Once the initial distribution of electrons is determined, the spatial and spectral distributions of energetic electrons in the coronal source and in the footpoints depends on the density of the ambiant medium $n$, the size of the acceleration region $d$ and the electron scattering mean free path $\lambda$ (see equation \ref{equation}). We search for the best set of parameters which can reproduce at the same time the spatial distribution of energetic electrons at 25 keV and the spectral distribution of energetic electrons in the footpoints (figures \ref{nvF} and \ref{spatial_nvf}). As described in section \ref{discussion_model}, each parameter affects both spatial and spectral distributions. The major effect is found for the width of the spatial distribution and the slope of the electron spectrum in the corona. As seen in table \ref{table_indexes}, although the slope of the energetic electrons in the footpoints never very far to the slope of the electron spectrum in the footpoints that will be produced in the standard case. 
The space of four parameters ($\dot{N}$, $n$, $\lambda$, $d$) was explored by producing predicted spatial and spectral distributions of energetic electrons that could be compared to the ones deduced from X-ray observations. A $\chi^2$ was computed for each set of parameter, as described in appendix \ref{fit}. The minimal $\chi^2$ was found for the following set of parameters: $\dot{N} = \left(4^{+0.5}_{-0.5}\right) \times 10^{34}$ s$^{-1}$, $n = \left(9.5^{+6.5}_{-2.5}\right) \times 10^{10}$cm$^{-3}$, $\lambda = \left(1.4^{+0.8}_{-0.4}\right) \times 10^8$ cm, $d = \left(5.5^{+0.7}_{-0.5}\right) \times 10^8$ cm. The uncertainties on the parameters represent the values for which the $\chi^2$ exceed the minimal $\chi^2$ by at least 5\% (see appendix \ref{fit}).
The modeled distributions are plotted in figures \ref{nvF} and \ref{spatial_nvf}. 
We can note that the slope of the looptop spectra is not well recovered by the model. On the other hand, the modeled spatial distribution does not seem to be as peaked that expected from the data. This is a consequence of a trade-off that happen during the fit to both spatial and spectral distributions. Nevertheless, the models fit the data with a density close to the density deduced from the observations, and a electron injection rate close to the electron rate deduced in the looptop source.

The spatial distribution of the energetic electron density is also computed with the set of parameters found above and compared with the values derived from the fit of the observations (see figure \ref{spatial_distribution_25}).  The cross-section of the loop $A_{LT}$ has been fixed to 26 Mm$^2$ (as this area has been used to calculate the electron density from the observations). 
In the different plots, the estimation of the error on the values of $\delta_{thick}$ and $ \dot{N}_{LT}$ is taken into account and is responsible for the error intervals around the distributions derived from the model and visible in figures \ref{spatial_nvf}, \ref{spatial_distribution_25} and \ref{spatial_distribution_60}.

%%%%%%%%%%%%%%%%%%%%%%%%%%%%%%%%%%%%%%%%%%%%%%%%%%%%%%%%%%%%%%%%%%%%%%%%%%%%%%%%%%%%%%%%%%%%%%%%%%%%%%%%%%%%%%%%%%%%%%%%%%%%%%%%%%%%%%%%%%%%%%%%%%%%%%%%%

\subsection{Comparison of radio observations with model predictions}
\label{radio_and_hxr}

The 2004 May 21 flare gyrosynchrotron emission has been studied by \cite{kuznetsov_and_kontar_2015}. The gyrosynchrotron emission is produced mostly by electrons of energies around 400 keV, and therefore the radio observations of the flare allows to study energetic electrons in a different energy domain than the X-ray analysis, X-ray emitting electrons being mostly in the 25-100 keV energy range.

\subsubsection{Comparison between X-ray and radio observations}

\label{obs_spatial_nb}

The spatial distributions of electrons at 25 keV (figure \ref{spatial_nvf}) and of the density of energetic electrons in the loop $n_b^{25}$ (figure \ref{spatial_distribution_25}), show that most of the energetic electrons with energy $E > 25$ keV are located in the looptop source. Moreover, an asymmetry is seen between the two footpoints. Both results are in agreement with the results obtained by \cite{kuznetsov_and_kontar_2015} who calculated the spatial distribution of the density of energetic electrons with energy $E > 60$ keV ($n_b^{60}$) from observations of the gyrosynchrotron emission \citep[see figure 7 in][and figure \ref{spatial_distribution_60} in the present paper]{kuznetsov_and_kontar_2015}. 
To compare with the results of \cite{kuznetsov_and_kontar_2015}, we estimate the number density of energetic electrons above 60 keV from the X-ray observations, using the relation $n_b^{60} \approx n_b^{25} \left(60/25\right)^{-\delta+1/2}$ with $\delta = 4.2$ : $n_b^{60} \approx 0.59 \times 10^6$ electrons cm$^{-3}$ in the corona.
This estimation of $n_b^{60}$ from X-ray producing electrons at the looptop is plotted as a cross in figure \ref{spatial_distribution_60}. 
It is about 2 times smaller than the value found by \cite{kuznetsov_and_kontar_2015} from radio observations (see figure \ref{spatial_distribution_60}). This difference could be explained if there is a break in the power-law spectrum of the energetic electrons with a smaller spectral index at higher energy or if the thin target approximation in the coronal source must be relaxed as suggested by the flatter spectrum of electrons in the coronal source derived in the model.

Independent of this quantitative comparison of relative numbers of electrons producing X-rays and radio emissions, we compare the relative spatial distributions of both X-ray emitting-electrons and radio emitting-electrons by comparing the ratio between the maximum electron density in the loop and the number density in the footpoints. 
The ratio between the number densities of energetic electrons at the looptop and in the footpoints $n_{b,LT}^{25}/n_{b,FP}^{25}$, from the X-ray measurements, are 1.6 and 3.8 for the first and the second footpoint respectively. 
Taking the values at the same distance in figure 7 in \cite{kuznetsov_and_kontar_2015}, the ratios $n_{b,LT}^{60}/n_{b,FP}^{60}$, averaged over the three times, are 7.7 and 9. We note that the ratio is much higher for the distribution deduced at energies above 60 keV from the gyrosynchrotron emission than for the one deduced at 25 keV from HXR observations. 
This implies that the X-ray emitting energetic electron spatial distribution is less strongly peaked than the spatial distribution deduced from microwave emissions, as seen in figure \ref{spatial_distribution_60} and that the high energy electrons responsible for gyrosynchrotron emissions are more confined in the corona than the lower energy ones. 
Based on the discussion of section \ref{discussion_model} and assuming that X-ray and radio emissions are produced by the same electrons injected and confined in the same loop (the values of $d$ and $n$ remains unchanged), the only way to produce a more spatially peaked distribution is to vary the scattering mean free path $\lambda$. Therefore, a second fit was performed on the distribution of the density of energetic electrons deduced from radio observations, the only free parameter being the scattering mean free path $\lambda$. The ambiant density and size of the acceleration region are kept as they were found by fitting the distributions deduced from X-ray observations. Figure 9 shows the modeled distribution with $\lambda = \left(1^{+4}_{-0.8}\right) \times 10^7$ cm, which produced the smallest $\chi^2$. Details of the fit are described in appendix \ref{fit}.

%%%%%%%%%%%%%%%%%%%%%%%%%%%%%%%%%%%%%%%%%%%%%%%%%%%%%%%%%%%%%%%%%%%%%%%%%%%%%%%%%%%%%%%%%%%%%%%%%%%%%%%%%%%%%%%%%%%%%%%%%%%%%%%%%%%%%%%%%%%%%%%%%%%%%%%%%%%%%%%%%%

\section{Discussion}
\label{discussion}

\label{section_discussion_models}

In this paper, we focused on the interpretation of our observations in the frame of the diffusive transport model described in \cite{kontar_et_al_2014}. We showed that the diffusive transport model of \cite{kontar_et_al_2014} can explain the observed trapping of energetic electrons in the corona. In particular, the model explains the electron spectrum in the footpoint, the hardening of the footpoint spectrum compared to the coronal spectrum, and the spatial distribution of energetic electrons along the loop. However, diffusive transport of energetic electrons in not the only mechanism that can explain electron trapping in the corona and we include a discussion about trapping with magnetic mirrors at the end of this section.

\subsection{Comparison with scattering mean free path in the interplanetary medium}

\begin{figure}
\begin{center}
\includegraphics[width=0.99\linewidth]{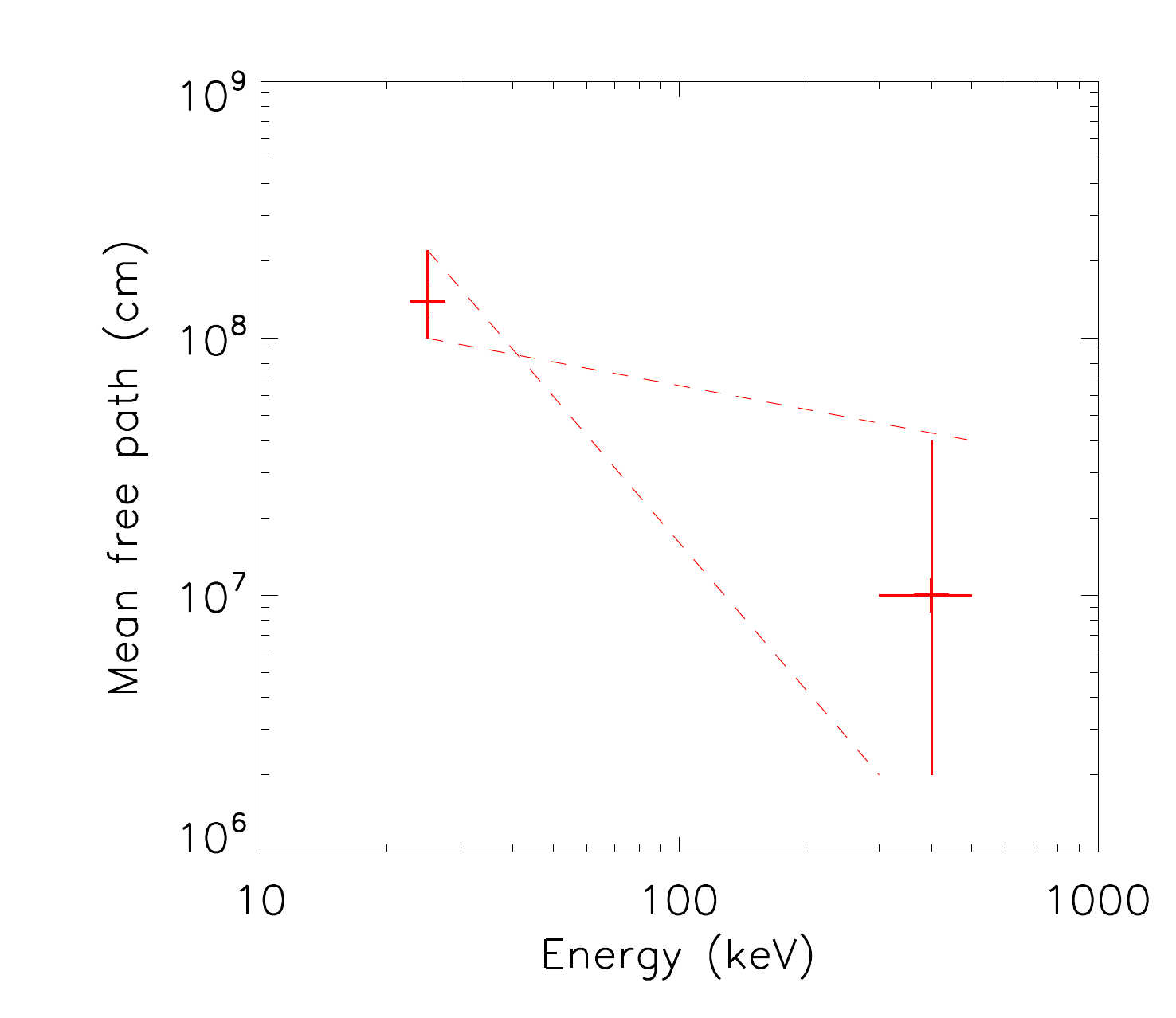}
\caption{Energy dependence of the scattering mean free path calculated with the diffusive transport model of \cite{kontar_et_al_2014}, with $A_{LT} = 26$ Mm$^2$, $d = 5.5$ Mm, $n = 9.5 \times 10^{10}$ cm$^{-3}$. The uncertainties on the values of the mean free path derived from the fit (see appendix \ref{fit}) and on the energy of the radio-emitting electrons are taken into account. The two most extreme slopes for the power law dependence of the mean free path are plotted.}
\label{consensus}
\end{center}
\end{figure}

The spatial distributions of X-ray emitting and radio emitting electrons can be reproduced in the context of the diffusive transport model of \cite{kontar_et_al_2014} only by assuming that the scattering mean free path of energetic electrons decreases with increasing electron energy, which explains why the trapping of energetic electrons in the corona is stronger at higher energies.
This conclusion is of course contradictory to the assumption of the model in which the scattering mean free path is constant with energy and shows that to completely study the behaviour of X-ray and radio emissions, a new model should be developed in which the scattering mean free path depends on energy. This is however out of the scope of the present paper.
We shall however discuss the result on the energy dependance of the scattering mean free path with respect with what is observed in the interplanetary medium. Several studies \cite[see e.g.][]{droge_2000,agueda_et_al_2014} have found for interplanetary electrons in range 0.1 - 1 MV a power law dependence of the electron mean free path on rigidity, with a negative power law index. We therefore also assume a power law dependence of the electron scattering mean free path with energy in the present study. We have only two data points, the first one derived from X-ray radiation above 25 keV, and the second one derived from radio observation, which is produced mostly by electrons at $400 \pm 100$ keV (private communication from A. Kuznetsov). 
The mean free paths calculated in this paper are plotted as a function of electron energy in figure \ref{consensus}. Given the uncertainty about the energy of radio-emitting electrons, and the uncertainty on the mean free path, we can calculate the slope of the power law in two limit cases: -1.9 and -0.3. The corresponding slopes for the mean free path dependence in rigidity are between -3.4 and -0.5. Although it is clear that the scattering mean free path is decreasing with increasing electron energy and rigidity, a large range of slopes are consistent with our data. It should be pointed out that, 5 out of 7 events studied by \cite{agueda_et_al_2014} have shown slopes for the rigidity dependence of the scattering mean free path of electrons in the interplanetary medium that could be consistent with our observations.

\subsection{Limitations and future improvements to the diffusive transport model}

\label{discussion_coronal_slope}

We note that the diffusive transport model predictions did not perfectly reproduce the observations in the details.
In particular, the predicted looptop source spectrum is flatter than the spectrum deduced from the X-ray analysis of the flare. This difference could be due to the fact that some effects are not taken into account in the diffusive transport equation that has been used in this paper, such as the effect of a converging magnetic field.
This discrepancy can also be explained by the fact that the thin target approximation might not be valid for the coronal source in this context. Indeed, the density calculated in the coronal source ($1.2 \pm 10^{11}$ cm$^{-3}$) is quite high and the source could be considered as a thick target for low energy electrons. Moreover, the diffusion of energetic electrons in the corona leads to enhanced time spent by the electrons in the target, where they loose more energy than assumed in the thin target approximation. However, assuming that the coronal X-ray source is a thick target in the spectral analysis do not improve the agreement between the data and the predictions of the diffusive transport model. The X-ray coronal source is most probably neither a thick nor a thin target, but is in between, with a density where none of these two approximations are completely valid.
Finally, the normalisation of the modeled distribution of the density of energetic electrons above 60 keV is not well recovered. This might indicate that fewer energetic electrons are accelerated at energies above 60 keV than expected (e.g. the injection electron rate decreases with energy), but it most probably due to the fact that the model produces a too-flat coronal spectrum and therefore overestimate the number of high-energy electrons in the loop.

We show that the diffusive transport can explain the observed spatial and spectral distributions in the X-ray and radio ranges, if the mean free path is energy dependent. The mean free path has been assumed to be constant in \cite{kontar_et_al_2014}: a further development of this model should include the energy dependance of the mean free path, as well as the relativistic effects, to allow a more precise comparison of the model prediction with combined X-ray and radio observations.

\subsection{Trapping with magnetic mirrors}

Trapping of energetic electrons in the coronal part of the loop can be explained by the effect of a converging magnetic field. In this event, the area of the section of the loop calculated (26 Mm$^2$) at the ends of the coronal source is larger than the area of the footpoints deduced from the X-ray observations (see table \ref{size_tab}); this observation is in favor of a magnetic convergence of the loop. If we considere a magnetic loss cone for the electron pitch angle distribution, the loss-cone angle $\alpha_{0}$ depends on the magnetic ratio $\sigma$, as described in the introduction.
The trapped fraction of the energetic electron distribution is deduced from X-ray observations and is $1- \frac{\dot{N}_{FP}}{\dot{N}_{LT}}$ \citep[see][for more details]{simoes_and_kontar_2013}.
\cite{simoes_and_kontar_2013} showed that in the case of an isotropic pitch-angle distribution, the trapped fraction of the energetic electron distribution is equal to $\mu_{0}$, the cosine of the losscone angle $\alpha_{0}$. We can therefore retrieve the value of $\sigma$ needed to explain the observed $\frac{\dot{N}_{LT}}{\dot{N}_{FP}}$ ratio in the case of an isotropic pitch-angle distribution : we found $\sigma \approx 1.4$, which is close to the values found by \cite{simoes_and_kontar_2013,aschwanden_et_al_2011,tomczak_and_ciborski_2007} and explains the observed ratio of cross-sections of the loop at the looptop and in the footpoints. This expected value $\sigma$ of the magnetic ratio can be compared to the magnetic ratio measured in the loop $\sigma_r$.

To estimate the magnetic ratio $\sigma_r$ of the coronal loop, we can use the magnetic extrapolation from \cite{kuznetsov_and_kontar_2015}. Note that in doing so, we assume that the HXR and gyrosynchrotron emissions are produced in a same magnetic loop, as mentioned in the introduction.
As it can be seen in Figure 6 in \cite{kuznetsov_and_kontar_2015}, at the looptop of the reconstructed magnetic loop, the magnetic field strength is $B_{LT} \approx 360$ G. 

Using the estimation of the source length seen in X-ray, we determined the value of the magnetic field at the supposed position of the mirrors (at each end of the observed coronal X-ray source) and found values of $430 \pm 30$ and $570 \pm 30$ G, leading to the following values of the magetic ratio: $\sigma_{r} \approx 1.2$ and $1.6$. This is consistent with the ratio of loop cross-sections at looptop and footpoints deduced from the X-ray images, $\approx 1.4$ and $1.5$ for the two footpoints. The magnetic ratio measured is therefore just enough to explain the ratio of electron rate $\frac{\dot{N}_{LT}}{\dot{N}_{FP}}$ deduced from the X-ray observations.
However, with this model, it is a priori not possible to explain why the trapping of energetic electrons is stronger at higher energies, and why a spectral hardening with a difference of 1 between electronic spectral slopes is observed between the looptop and footpoint sources.
We can also note that \cite{kuznetsov_and_kontar_2015} showed a shift between the centroid of the gyrosynchroton source and the top of the magnetic loop where the magnetic field is minimal. In the case of electron trapping due to magnetic mirroring, we expect to have a maximum emission where the magnetic field is minimum.

%%%%%%%%%%%%%%%%%%%%%%
% *** CONCLUSION *** %
%%%%%%%%%%%%%%%%%%%%%%

\section{Summary and conclusion}
\label{conclusion}

The summary of our observations is the following:
\begin{enumerate}
\item The difference between the footpoint and looptop spectral indexes is about 1, which suggests that a mechanism is hardening the electron spectrum during the transport. This can be explained by trapping of energetic electrons in the corona.
\item The ratio of the looptop and footpoint electron rate above 25 keV, $\frac{\dot{N}_{LT}}{\dot{N}_{FP}}$, has a value of 2.2, suggesting that part of the energetic electrons are trapped in the coronal part of the loop.
\item The spatial distribution of HXR-emitting electrons is peaked near the looptop, but less peaked than the spatial distribution of microwave-emitting electrons since the ratio of energetic electron density between the looptop and the footpoints is more than two time higher for radio-emitting electrons above 60 keV than X-ray emitting electrons above 25 keV.
\item The spectral and spatial distribution of energetic electrons, deduced from both X-ray and radio observations, can be explained by a diffusive transport model of \cite{kontar_et_al_2014}, with a mean free path decreasing with increasing electron energy. 
\item The mean free path for electron energies between 25 and 100 keV is of the order of $1.4 \times 10^8$ cm, which is smaller than the length of the loop. These values are comparable to values found by \cite{kontar_et_al_2014}. The mean free path is also smaller than the size of the acceleration region calculated with the model ($5.5 \times 10^8$ cm), which suggests that electrons can potentially be accelerated for a longer time.
\item The scattering mean free path for electron energies around 400 keV ($10^7$ cm) is significantly smaller than the mean free path estimated at lower energies. Similar dependence of the scattering mean free path over electron energies has been found in the case of interplanetary electron transport, in the same range of electron energy. We note that the potential slopes of the energy dependence of the scattering mean free path in the solar corona are in agreement with some of the slopes observed for interplanetary electrons.
\end{enumerate}

Trapping due to magnetic mirroring is not known to be energy dependent and this mechanism cannot fully explain our observations.

The diffusive transport model enable the reproduction of our different observations, such as the spectral slope in the footpoints, some spectral hardening between the looptop and the footpoints, and the spatial distributions of electron density deduced from both X-ray and radio observations.

Imaging spectroscopy in HXR is a powerful tool to study electron transport during solar flares. This study should encourage the development of predictions on the spatial distribution of electrons and on the evolution of the spectral index of non-thermal electron energy distribution by the various transport models.
The simultaneous observation of non-thermal X-ray sources in both the coronal part of the loop and its footpoints is rare due to the facts that footpoints sources are usually brighter than coronal sources and that indirect imaging intruments such as RHESSI have a limited dynamic range. Instruments using focusing optics for hard X-ray imaging (such as a FOXSI spacecraft) would therefore provide very useful observations of faint non-thermal coronal sources in the presence of bright footpoints and therefore add interesting cases in which to study electron transport during flares.

Finally, this study shows that the combination of X-ray and radio diagnotics for energetic electrons in closed loops during flare enable to study the energy dependence of transport properties in the solar corona, such as the scattering mean free path. Such observations could constrain in some extend some properties of the turbulence spectrum in the solar corona.

\begin{acknowledgements}
We thank the RHESSI team for producing free access to data, Alexey Kuznetsov for his help with the radio data, and our referee for his useful comments. Sophie Musset acknowledges the CNES and the LABEX ESEP (N$^{\circ}$ 2011-LABX-030) for the PhD funding, and thanks the French State and the ANR for their support through the \textquotedblleft Investissements d'avenir\textquotedblright \ programm in the PSL$\ast$ initiative (convention N$^{\circ}$ ANR-10-IDEX-0001-02), as well as the Programme National Soleil-Terre (PNST). EPK was supported by a STFC consolidated grant ST/L000741/1.
\end{acknowledgements}

\appendix

\section{CLEAN beam factor}
\label{app_bf}

The CLEAN algorithm is an iterative algorithm based on the assumption that the X-ray image is well represented by a superposition of point sources convolved with the point spread function (PSF) of the instrument \citep[see e.g.][]{rhessi_imaging}. The CLEAN algorithm developed for the RHESSI image analysis has one parameter called 'beam factor', which represents the effective resolution of the subcollimators used to reconstruct the image.

In this paper, the value for the beam factor was chosen to have CLEAN images as close as possible to images reconstructed with the visibility forward fit VISFF (see \citealt{schmahl_et_al_2007} for the definition of visibilities and \citealt{xu_et_al_2008} for examples of application) and the PIXON \citep{metcalf_et_al_1996,rhessi_imaging} algorithms. This is a standard procedure to ensure that CLEAN agrees with other algorithms for the image reconstruction \citep[see e.g.][]{dennis_and_pernak_2009,kontar_et_al_2010}.

 The determination of the best value of the beam factor for the image reconstruction has an important impact on the X-ray source size determination on CLEAN images.
For example, when using the default value of the beam factor 1, the measured sizes are roughly 1.5 times greater than the sizes estimated on CLEAN images with a beam factor of 1.7 or on a PIXON image.

\section{X-ray production in thin- and thick-targets}
\label{app_eqn}

The bremsstrahlung photon flux at energy $\epsilon$, $I(\epsilon)$, produced by an energetic electron flux density distribution $F(E, \vec{r})$ (electrons/cm$^{2}$/s/keV) in an emitting source (a target) of plasma density $n$ and volume $V$ is expressed as:
\begin{equation}
I(\epsilon) = \frac{1}{4 \pi R^{2}} \int_{V} \int_{\epsilon}^{\infty} n(\vec{r}) F(E,\vec{r}) Q(\epsilon, E) dE dV
\label{x-ray_spectrum}
\end{equation}
Where $Q(\epsilon,E)$ is the differential bremsstrahlung cross-section, and the integration is done over the target volume and all contributing electron energies, which are all electron energies above the photon energy $\epsilon$.

We can see that the X-ray spectrum $I(\epsilon)$ is linked to both the energetic electron distribution and the ambiant plasma properties (density and volume of the target). 

For spectral observations, we deal with a spatially-integrated form of equation \ref{x-ray_spectrum}:
\begin{equation}
I(\epsilon) = \frac{1}{4 \pi R^{2}} \int_{\epsilon}^{\infty} \left\langle \bar{n} V \bar{F}(E) \right\rangle Q(\epsilon, E) dE 
\label{int_x-ray_spectrum}
\end{equation}
where $\bar{n} = (1/V) \int_{V} n(\vec{r}) dV$ and $\bar{F}(E)$ (electrons/cm$^{2}$/s/keV) is the mean electron flux distribution, i.e. the plasma-density-weighted, target-averaged electron flux density distribution \citep{brown_et_al_2003,kontar_et_al_2011_review,holman_et_al_2011}, defined as:
\begin{equation}
\bar{F}(E) = \frac{1}{\bar{n}V} \int_{V} n(\vec{r}) F(E,\vec{r}) dV
\label{}
\end{equation}

Since the quantity $\bar{n} V$ is dimensionless, the units of $\left\langle \bar{n} V \bar{F}(E) \right\rangle$ are the same as those of the electron flux (electrons/cm$^{2}$/s/keV). $\left\langle \bar{n} V \bar{F}(E)\right\rangle$ is a quantity which can be retrieved from the X-ray spectrum $I(\epsilon)$ without any model assumption and therefore, is the quantity derived during spectroscopic diagnosics of the X-ray emission.
To retrieve the product $\left\langle \bar{n}V\bar{F} \right\rangle$, in principle, we only need to know the bremsstrahlung cross-section $Q(\epsilon, E)$.

In our study, we were particularly interested by the number density of energetic electrons with energy $E > E_{min}$, $n_{b}^{E_{min}}$ (in electrons cm$^{-3}$), which is defined as:
\begin{equation}
n_{b}^{E_{min}} \equiv \int^{\infty}_{E_{min}} \frac{F(E)}{v} dE
\label{density_def_2}
\end{equation}
where $v$ is the velocity of the electrons. It can also be expressed as:
\begin{equation}
n_{b}^{E_{min}} \equiv \int^{\infty}_{E_{min}} \frac{ \left\langle \bar{n}V\bar{F}(E) \right\rangle}{\bar{n}Vv} dE
\label{eqn_nb}
\end{equation}

We distinguish two approximations, the thin-target and the thick-target models. In the thin target model, energetic electrons lose only a small fraction of their energy while they pass through the target, whereas in the thick target model, energetic electrons lose all their supra-thermal energy in the target. 

In the following, we describe how the product $\left\langle \bar{n} V \bar{F}\right\rangle$ is expressed in the thin- and thick-target models in OSPEX and how we estimate the energetic electron number density $n_{b}$ (cm$^{-3}$).

\subsection{Thin target model}

We assume a power-law distribution for the electron mean spectrum: $\bar{F}(E) \propto E^{- \delta_{thin}}$. In OSPEX, the proportionality constant is defined such as we can write the spatially integrated density weighted mean flux spectrum $ \left\langle \bar{n}V\bar{F}(E) \right\rangle$ (in electrons s$^{-1}$ cm$^{-2}$ keV$^{-1}$) as:
\begin{equation}
\left\langle \bar{n}V\bar{F}(E) \right\rangle = \left\langle \bar{n}V\bar{F}_{0} \right\rangle \frac{\delta_{thin}-1}{E_{0}} \left(\frac{E}{E_{0}}\right)^{-\delta_{thin}},  E > E_{0}
\label{mean_flux_thin}
\end{equation}
where $\delta_{thin}$ and $\left\langle \bar{n}V\bar{F}_{0} \right\rangle = \left( \int^{\infty}_{E_{0}} \left\langle \bar{n}V\bar{F}(E) \right\rangle dE \right)$ are the spectral index and the normalisation factor given by the spectral analysis (see table \ref{spectral_analysis_tab}).

Equation \ref{eqn_nb} and can be integrated over E, using equation \ref{mean_flux_thin} to obtain:
\begin{equation}
n_{b}^{E_{min}} = \frac{\left\langle \bar{n}V\bar{F}_{0} \right\rangle}{\bar{n}V} \frac{\delta_{thin}-1}{\delta_{thin}-1/2} E_{min}^{-1/2} \sqrt{m/2} \left(\frac{E_{0}}{E_{min}} \right)^{\delta_{thin}-1}
\label{nb_thin}
\end{equation}
where $m$ is the electron mass (in keV$/$c$^{2}$). 

\subsection{Thick target model}

In the thick target model, energetic electrons lose all their supra-thermal energy through efficient collisions. Therefore, the energetic electron spectrum $\bar{F}$ is different from the injected electrons spectrum $F_{0}$. In fact, we need to integrate the injection spectrum over all energies in the X-ray emitting source.

Therefore, the number of photons of energy between $\epsilon$ and $\epsilon + \delta \epsilon$ produced by an electron of initial energy $E_{0}$ is:
\begin{equation}
\nu (\epsilon, E_{0}) = \int_{t=0}^{t_{F}} n(\vec{r}) Q(\epsilon, E(t)) v(t) dt
\label{}
\end{equation}
where $t_{F}$ is the time at which all energetic electrons have been thermalized. 
Since energetic electrons are losing energy at a rate $dE/dt$, the time integration can be replaced by an integration over energy:
\begin{equation}
\nu (\epsilon, E_{0}) = \int_{\epsilon}^{E_{0}} \frac{n(\vec{r}) Q(\epsilon, E) v(E)}{\left|dE/dt\right|} dE
\label{nbr_photons}
\end{equation}

Energetic electrons lose their energy by Coulomb collisions with the electrons of the ambient plasma, and in that case the energy loss rate is expressed as:
\begin{equation}
dE/dt = -(K/E) n(\vec{r}) v(E)
\label{energy_loss_rate}
\end{equation}
where $K = 2 \pi e^{4} \Lambda$, with $\Lambda$ is the Coulomb logarithm, $e$ is the electron charge, $n$ the density of the plasma, and $v$ is the speed of the energetic electron.

If we consider the injected electron spectrum $F_{0}(E_{0})$, the X-ray spectrum can be express as:
\begin{equation}
I(\epsilon) = \frac{A}{4 \pi R^{2}} \int_{E_{0}=\epsilon}^{\infty} F_{0}(E_{0}) \nu (\epsilon, E_{0}) dE_{0}
\label{int2_x-ray_spectrum}
\end{equation}
where $A$ is the area of the thick target source.

Using equation \ref{energy_loss_rate} in equation \ref{nbr_photons}, we can rewrite equation \ref{int2_x-ray_spectrum} in the following way:
\begin{equation}
I(\epsilon) = \frac{A}{4 \pi R^{2}} \frac{1}{K} \int_{E_{0}=\epsilon}^{\infty} F_{0}(E_{0}) \int_{E=\epsilon}^{\infty} E Q(\epsilon, E) dE dE_{0}
\label{}
\end{equation}
and by changing the integration order, and comparing with equation \ref{int_x-ray_spectrum}:
\begin{equation}
\left\langle \bar{n}V\bar{F}(E) \right\rangle = A \frac{E}{K} \int_{E_{0}=E}^{\infty} F_{0}(E_{0}) dE_{0}
\label{eq_thick_1}
\end{equation}

Once again, we assume the injection spectrum to have a power-law dependence in energy, $F_{0} \propto E_{0}^{-\delta_{thick}}$. In OSPEX, the injection spectrum $F_{0}(E)$ (electrons$/$sec$/$cm$^{2}/$keV) has the following form:
\begin{equation}
F_{0}(E) = \frac{\dot{N}}{A} \frac{\delta_{thick}-1}{E_{0}} \left( \frac{E}{E_{0}} \right)^{-\delta_{thick}}, E > E_{0}
\end{equation}
where $\dot{N}$ is the injection electron rate (in electrons s$^{-1}$), and $\delta_{FP}$ is the spectral index.

After integration of equation \ref{eq_thick_1}, the spatially integrated density weigthed mean flux spectrum is:
\begin{equation}
\left\langle \bar{n}V\bar{F}(E) \right\rangle = \frac{\dot{N}}{K} E_{0} \left( \frac{E}{E_{0}} \right)^{-\delta_{thick} + 2}
\label{mean_flux_thick}
\end{equation}

Equation \ref{eqn_nb} is also valid for the thick target model. Using equation \ref{mean_flux_thick} and after integration, the density of energetic electrons, in electrons$/$cm$^{3}$, in the thick target, is:
\begin{equation}
n_{b}^{E_{min}} = \frac{\dot{N}}{K} \frac{\sqrt{m/2}}{\bar{n}V} \frac{E_{min}^{3/2}}{\delta_{thick}-5/2}  \left( \frac{E_{0}}{E_{min}} \right)^{\delta_{thick} -1}
\label{nb_thick}
\end{equation}

\section{Model fitting}
\label{fit}

The first fit of the model to the data was performed using the X-ray observations. The electron mean spectra for the coronal source and one footpoint, as well as the spatial distribution of electrons at 25 keV, are modeled and compared to the same distributions deduced from the X-ray observations. These distributions are visible in figure \ref{nvF} and \ref{spatial_nvf}. The $\chi^2$ is calculated by comparing the looptop spectra between 22 keV and 39 keV, where the observed spectra are mostly non-thermal; by comparing the footpoint spectra between 24 and 100 keV ; and by comparing the spatial distribution a the three data points deduced from the observations. The errors on the observations are derived from the errors found on the free parameters in the spectral analysis (see table \ref{spectral_analysis_tab}).

\begin{figure*}
\begin{center}
\includegraphics[width=0.95\linewidth]{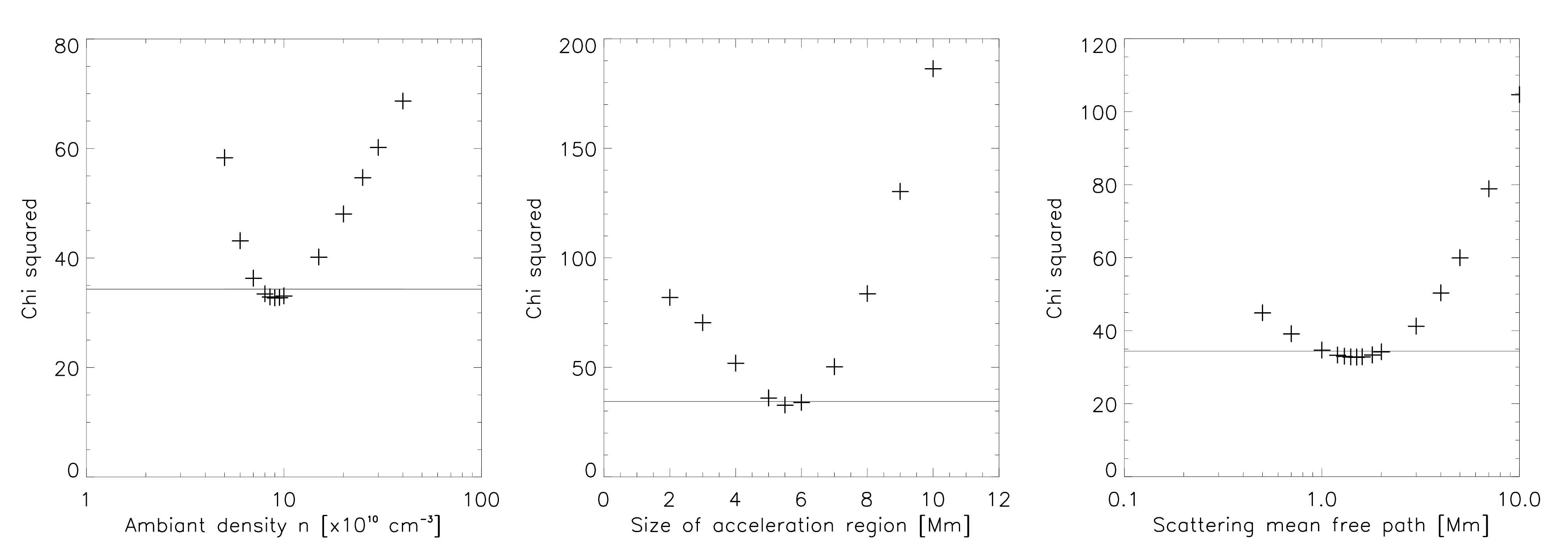}
\caption{Evolution of the values of the $\chi^2$ with the free parameters in the model. Left: evolution of $\chi^2$ with ambiant density $n$, with $\lambda = 1.4 \times 10^8$ cm and $d = 5.5$ Mm. Middle: evolution of $\chi^2$ with the size of the acceleration region $d$, with $\lambda = 1.4 \times 10^8$ cm and $n = 9.5 \times 10^{10} $ cm$^{-3}$. Right: evolution of $\chi^2$ with the scattering mean free path, with $d = 5.5$ Mm and $n = 9.5 \times 10^{10} $ cm$^{-3}$. The horizontal line marks the limit of 5\% of the minimal $\chi^2$ value that as been used to determine uncertainties on the best values for the model free parameters.}
\label{figure_evolution_chi2}
\end{center}
\end{figure*}

The evolution of the $\chi^2$ in regards to the free parameters ($n$, $\lambda$, $d$) is displayed in figure \ref{figure_evolution_chi2}. To provide uncertainties on the values of those parameters, we looked at the values for which the $\chi^2$ was 5\% larger than its minimum. The resulting density is between $7 \times 10^{10}$ and  $1.6 \times 10^{11}$ cm$^{-3}$, the size of the acceleration region is between $5$ and $6.2$ Mm and the scattering mean free path is between $1 \times 10^{8}$ and $2.2 \times 10^{8}$ cm.

We note that the final value of the $\chi^2$ is quite big, which is due in particular to the fact that the spectral slope of the coronal spectrum is not well recovered.

The fit of the model to the spatial distribution of the density of energetic electrons above 60 keV deduced from radio observations was performed by comparing the modeled distributions on artificially created data points between -17 and +17 Mm, spaced of 0.5 Mm each. The error on the distribution deduced from observations was set to 10\% of the value, since this is the maximum error on that distribution according to \cite{kuznetsov_and_kontar_2015}. The range of values of the scattering mean free path that lead to the best fit within 5\% of the minimum $\chi^2$ is $2 \times 10^6$ to $4 \times 10^7$ cm.

\bibliographystyle{aa} % style aa.bst
\bibliography{zmabiblio} % your references Yourfile.bib

\end{document}